\documentclass[onecollarge,final,natbib]{svjour2}

\usepackage[dvips]{graphicx}
\usepackage{exscale}
\usepackage{amsmath}
\usepackage{amsfonts}
\usepackage{amssymb}

\newcommand{\mathtext}[1]{\hspace{0.5em}\mbox{#1}\hspace{0.5em}}
\newcommand{\grd}[1]{\ensuremath{\boldsymbol \nabla} #1}
\renewcommand{\div}[1]{\ensuremath{\boldsymbol \nabla\!\cdot\!} #1}
\newcommand{\crl}[1]{\ensuremath{\boldsymbol \nabla\!\times\!} #1}
\newcommand{\ngrd}[1]{(\vec{n}\cdot\ensuremath{\boldsymbol{\nabla}}) #1}
\newcommand{\sdot}{\ensuremath{\!\cdot\!}}
\newcommand{\crss}{\ensuremath{\!\times\!}}
\newcommand{\bra}{\ensuremath{<\!}}
\newcommand{\ket}{\ensuremath{\!>}}
\journalname{Solar Physics}
  
\begin{document}

\title{\bf Nonlinear force-free magnetic field extrapolations:
comparison of the Grad-Rubin and Wheatland-Sturrock-Roumeliotis algorithm}

\author{Bernd Inhester \and  Thomas Wiegelmann}

\institute{Max-Planck-Institute f\"ur Sonnensystemforschung
37191 Katlenburg-Lindau, Germany}

\titlerunning{Nonlinear force-free magnetic field extrapolations}
\authorrunning{B. Inhester et al.}

\date{received: 12 December 2005,~
      accepted: 20 January 2006\\
      Solar Physics, Volume 235, Issue 1-2, pp. 201-221,~
      doi:10.1007/s11207-006-0065-x}

\maketitle

\begin{abstract}
We compare the performance of two alternative algorithms which aim to
construct a force-free magnetic field given suitable boundary
conditions. For this comparison, we have implemented both algorithms
on the same finite element grid which uses Whitney forms to describe
the fields within the grid cells. The additional use of conjugate
gradient and multigrid iterations result in quite effective codes.

The Grad-Rubin and Wheatland-Sturrock-Roumeliotis algorithms
both perform well for the reconstruction of a known analytic force-free field.
For more arbitrary boundary conditions the Wheatland-Sturrock-Roumeliotis 
approach has some difficulties because it requires overdetermined
boundary information which may include inconsistencies.
The Grad-Rubin code on the other hand loses convergence for strong
current densities.
For the example we have investigated, however, the maximum possible
current density seems to be not far from the limit beyond which a
force free field cannot exist anymore for a given normal magnetic
field intensity on the boundary.
\end{abstract}
  
\section{Introduction}
\label{sec:Intro}

With the advent of vector magnetographs which measure the line-of-sight
component of the photospheric magnetic field and, except for a 180$^{\circ}$
ambiguity, also its component normal to the line-of-sight, the interest in
extrapolating these measurements into the corona have grown enormously.

The line-of-sight component of the photospheric magnetic field has
been observed for decades now but these measurements alone supply only
boundary information at best sufficient for a Laplace field model of
the coronal magnetic field. The vector magnetograph observations now
available considerably constrain the photospheric horizontal field and
allow estimates also of the coronal current density. As a consequence
much more realistic coronal field models can be based on these
observations.

Since the magnetic field, at least in the lower corona, completely
dominates the plasma forces, a ``force-free'' approximation of the
field for stationary situations seems to be a tolerable assumption.
``Force'' in this context means the magnetic Lorentz force
$\vec{j}\crss\vec{B}$. All other MHD forces like gravity, pressure,
etc. are neglected because they are about three orders of magnitude
smaller than $jB$. Hence, the current and magnetic field vectors
should be aligned to better than half a degree.

These simplifications accepted, 
the magnetic field in some domain $V$ of the corona may be
described by
\begin{equation}
  \div{\vec{B}}=0 \;;\quad \crl{\vec{B}}=\vec{j} \;;\quad
  \vec{j}\crss\vec{B}=0\,.
\label{FF}\end{equation}
The alignment of current and magnetic field causes the problem to be
nonlinear, hence the questions which boundary information is to be
supplied and how to solve (\ref{FF}) are by no means trivial.

Boundary conditions which seem necessary and sufficient for (\ref{FF})
are \citep{Boulmezaoud:Amari:2000}
\begin{equation}
  \vec{n}\sdot\vec{B} \mathtext{on} \partial V \;;\quad
  \vec{n}\sdot\crl{\vec{B}} \mathtext{on} (\partial V)^-
  \mathtext{or} (\partial V)^+\,,
\label{BC}\end{equation}
where $(\partial V)^{\pm}$ is that part of the surface of $V$ where
$\vec{n}\sdot\vec{B}$ is either $>$ 0 or $<$ 0.
But not all the boundary values which comply with (\ref{BC}) are allowed.
The field line twist which can be stationarily maintained, i.e.,
$\vec{n}\sdot\crl{\vec{B}}$ on $\partial V$, is limited by the
field energy sustained from the exterior, i.e., by 
$\vec{n}\sdot\vec{B}$ (see section \ref{results:TwistedLoop}).

Quite some effort has gone into attempts to solve (\ref{FF}) for given
boundary values either in the form (\ref{BC}) or differently. Among
the most promising schemes are those suggested a long time ago by Grad
and Rubin (\citeyear{Grad:Rubin:1958}, abbreviated GR) and more recently
by Wheatland, Sturrock, and Roumeliotis (\citeyear{Wheatland:etal:2000},
abbreviated WSR). The GR code has first been implemented by Sakurai
(\citeyear{Sakurai:1981}) and further developed by Amari et al.
(\citeyear{Amari:etal:1999a}), R\'egnier et al. (\citeyear{Regnier:etal:2002})
and Wheatland (\citeyear{Wheatland:2004}).
The WSR scheme has been extended by Wiegelmann (\citeyear{Wiegelmann:2004})
and Wiegelmann and Inhester (\citeyear{Wiegelmann:Inhester:2003}).
The development still is an active area of research. A comparison
and description of codes in use and other alternative approaches can
be found in Schrijver et al. (\citeyear{Schrijver:etal:2006}).

A difficulty for a thorough comparison of the various schemes,
however, is the fact that they are often implemented very heterogeneously
and also include many different details which could easily speed up or
slow down their performance and sometimes may even obscure the basic
advantages or disadvantages of an approach.

In order the make the two schemes comparable, we apply them to the same
problem defined on exactly identical grids. Error norms to measure the
performance are exactly the same.
The special grid which we use is explained in section~\ref{sec:Grid}.
We are convinced that it has many advantages for electromagnetic problems
like (\ref{FF}). 

The WSR and the GR algorithms have been extensively described
elsewhere, so that in sections \ref{sec:Wheatland} and \ref{sec:GradRubin}
we restrict ourselves to the basics and rather emphasize some or the
details of our implementation.
The numerical results obtained for two different examples are presented
and discussed in the final parts \ref{sec:Results} and \ref{sec:Discussion}. 

\section{The grid}
\label{sec:Grid}

For the problem we want to investigate the choice of the grid and the
representation of the fields is very crucial. We found most suitable
for our purposes a finite element grid which allows to transform
standard vector analysis consistently into discrete space.
It is related to finite difference grids with staggered field
components and similar grids based on Yee's scheme \citep{Yee:1966}.  
In part of the mathematical literature these special finite elements
are called discrete Whitney forms \citep{Bossavit:1988} because they have very
much in common with continuous differential forms. In fact, some of the
finite elements have been known for a long time, however, the way they
are related among each other by differentiation operations and to
their dual space analogues is relatively new and a matter of current
research \citep{Hiptmair:2001,Gradinaru:2002}.
The elements are particularly suited for a numerical treatment of
electromagnetic fields \citep{Teixeira:2001}.

We here use the elements in the most simple form to lowest order and on a
regular cubic grid which spans our computational domain, a square box
$V$ = $[0,1]\times[0,1]\times[0,1]$.
The $n$$\times$$n$$\times$$n$ grid cells each have a size $h$ = $1/n$.
The cell vertices are located at $ih$, $i=0,n$; the cell centres
are at $(i-0.5)h$, $i=1,n$.

\begin{figure}
\hspace*{\fill}
\begin{picture}(14.5,4.2)
  \put( 0.,4.2){\includegraphics[bb=70 53 220 770,clip,width=3cm,angle=-90]
            {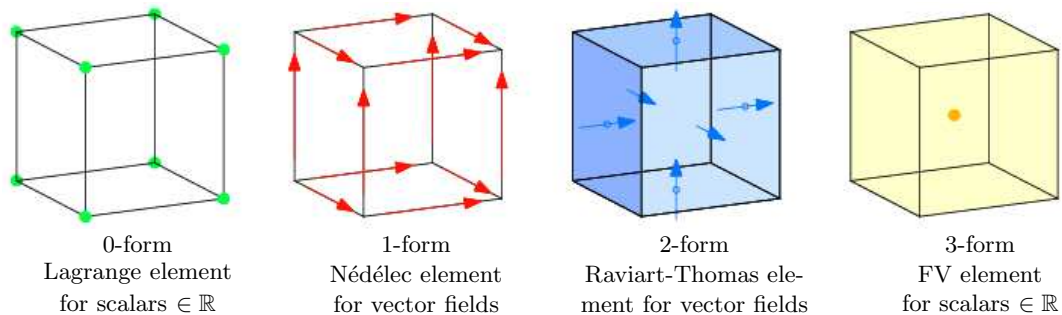}}
  \put(1.8,1.){\makebox(0,0){\small 0-form}} 
  \put(1.8,0.6){\makebox(0,0){\small Lagrange element}}
  \put(1.8,0.2){\makebox(0,0){\small for scalars $\in \mathbb{R}$}}
  
  \put(5.5,1.){\makebox(0,0){\small 1-form}} 
  \put(5.5,0.6){\makebox(0,0){\small N\'ed\'elec element}}
  \put(5.5,0.2){\makebox(0,0){\small for vector fields}}
  
  \put(9.2,1.){\makebox(0,0){\small 2-form}} 
  \put(9.2,0.6){\makebox(0,0){\small Raviart-Thomas ele-}}
  \put(9.2,0.2){\makebox(0,0){\small ment for vector fields}}
  
  \put(13,1.){\makebox(0,0){\small 3-form}} 
  \put(13,0.6){\makebox(0,0){\small FV element}}
  \put(13,0.2){\makebox(0,0){\small for scalars $\in \mathbb{R}$}}
\end{picture}
\hspace*{\fill}
\caption{Lowest order Whitney forms for a square grid cell.}
\label{fig:Whitney-elements}\end{figure}

In Figure~\ref{fig:Whitney-elements} we present the four types of finite
elements which we use here. Each form a Hilbert space of either scalar
or vector valued piecewise first or zero order polynomials determined
by a set of parameters which are representative averages of the field
to be described. The respective averaging areas are indicated in colour.

Lagrangian elements (0-forms) have as parameters the function values at
the 8 vertices of a cell. The element function is linear inside the grid cell
such that the correct values are met at the vertices.

The N\'ed\'elec elements (1-forms, \citeauthor{Nedelec:1986}
\citeyear{Nedelec:1986}) are a discrete representation of a vector
field. Its element parameters are the averages of a field component
over a cell edge along the respective component direction. The element
function is constant along the edge and varies linearly in transverse
direction matching the right values at the four cell edges of the
given direction.

The Raviart-Thomas element (2-forms, \citeauthor{Raviart:Thomas:1977}
\citeyear{Raviart:Thomas:1977}) has as element parameters the face
averages of the field component normal to the face. The element
function varies linearly in this normal direction and is constant
across the face plane.

The last element we need is a finite volume element (3-form) for a scalar
function approximation. It has only a single parameter per grid cell
which represents the average of the scalar over the entire cell.

Every vector differential operation transforms an $n$-form element in
a natural way into a $n\!+\!1$-form element:
\\
  \hspace*{\fill}
\begin{picture}(11,1.0)
  \put(  1,0.5){\makebox(0,0){0-form}}
  \put(2.5,0.7){\makebox(0,0){grad}} 
  \put(2.5,0.6){\makebox(0,0){\vector(1,0){1}}} 
  \put(  4,0.5){\makebox(0,0){1-form}} 
  \put(5.5,0.7){\makebox(0,0){curl}} 
  \put(5.5,0.6){\makebox(0,0){\vector(1,0){1}}} 
  \put(  7,0.5){\makebox(0,0){2-form}} 
  \put(8.5,0.7){\makebox(0,0){div}} 
  \put(8.5,0.6){\makebox(0,0){\vector(1,0){1}}} 
  \put( 10,0.5){\makebox(0,0){3-form}} 
\end{picture}
\hspace*{\fill} \\
As for continuous differential forms, a double differentiation gives
exactly a zero field and insures thus that curl$\circ$grad and div$\circ$curl
vanish identically also for the discrete forms. This vanishing of
double differentiation is not a consequence of the precision with
which we approximate the differentiation of discrete forms but is
due to the fact that the boundary of a boundary is an empty set,
hence is a consequence of geometry alone \citep{Janich:2001}.
It is therefore not surprising that this rule holds also
for discrete forms.

\begin{figure}
  \hspace*{\fill}
\begin{picture}(11.5,4.2)
  \put( 3,4.2){\includegraphics[bb=111 61 333 311,clip,width=4.5cm,angle=-90]
           {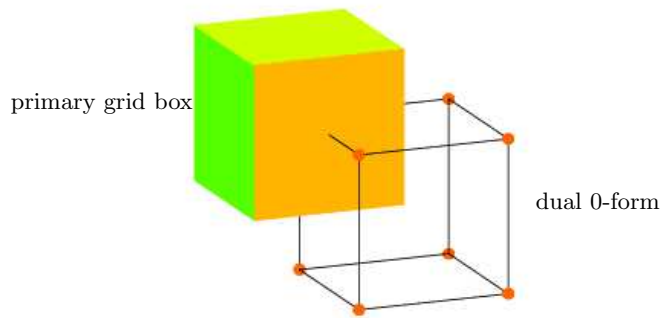}}
  \put(1.5,2.8){\makebox(1,0){primary grid box}} 
  \put(8.1,1.5){\makebox(1,0){dual 0-form}} 
\end{picture}
  \hspace*{\fill}
  \caption{Dual grid cell in relation to a primary (coloured) grid cell.}
\label{fig:DualGrid}\end{figure}

In order to allow for Laplacians, the alternative combination of vector
differential operators, we need to introduce the dual grid as opposed
to the primary grid described above. The dual grid is in case of the
regular square primary grid again a square grid but shifted by half a
grid size in each axis direction such that vertices of the dual grid
are located at the cell centres of the primary grid (see
Figure~\ref{fig:DualGrid}). Hence there is a natural association between
primary 3-forms and dual 0-forms. This association also holds for the
other forms, between the $n$-forms of the dual grid and primary
$(3-n)$-forms. For general grids this association is manifested by the
Hodge (or $\star$) transform \citep{Hiptmair:2001}. For the square grid
we use here, the Hodge transform is simple because the finite element
parameters of the dual form are the same as the corresponding parameters
of the primary element (except for domain boundary effects).
Note, however, that while the finite element parameters remain unaltered
under this transformation, their interpretation and their functional
representation changes.

Of course, the forms of the dual grid are connected among each other
by differentiations in the same way as for the primary grid. The final
pattern of forms together with the mappings among them yields\\
  \hspace*{\fill}
\begin{picture}(14.5,3)
  \put( 1.6,2.3){\makebox(0,0){primary grid:}}
  \put( 2,0.5){\makebox(0,0){dual grid:}}
  
  \put(  4,2.3){\makebox(0,0){0-form}}
  \put(5.5,2.5){\makebox(0,0){grad}} 
  \put(5.5,2.4){\makebox(0,0){\vector(1,0){1}}} 
  \put(  7,2.3){\makebox(0,0){1-form}} 
  \put(8.5,2.5){\makebox(0,0){curl}} 
  \put(8.5,2.4){\makebox(0,0){\vector(1,0){1}}} 
  \put( 10,2.3){\makebox(0,0){2-form}} 
  \put(11.5,2.5){\makebox(0,0){div}} 
  \put(11.5,2.4){\makebox(0,0){\vector(1,0){1}}} 
  \put( 13,2.3){\makebox(0,0){3-form}}
  
  \put(  4.1,1.4){\makebox(0,0){\vector(0,-1){1.2}}}
  \put(  7.1,1.4){\makebox(0,0){\vector(0,-1){1.2}}} 
  \put( 10.1,1.4){\makebox(0,0){\vector(0,-1){1.2}}} 
  \put( 13.1,1.4){\makebox(0,0){\vector(0,-1){1.2}}}
  \put(  4,1.4){\makebox(0,0){\vector(0,1){1.2}}}
  \put(  7,1.4){\makebox(0,0){\vector(0,1){1.2}}} 
  \put( 10,1.4){\makebox(0,0){\vector(0,1){1.2}}} 
  \put( 13,1.4){\makebox(0,0){\vector(0,1){1.2}}}
  \put(  4.3,1.4){\makebox(0,0){$\star$}}
  \put(  7.3,1.4){\makebox(0,0){$\star$}} 
  \put( 10.3,1.4){\makebox(0,0){$\star$}} 
  \put( 13.3,1.4){\makebox(0,0){$\star$}}

  \put(  4,0.5){\makebox(0,0){3-form}}
  \put(5.5,0.7){\makebox(0,0){div}} 
  \put(5.5,0.4){\makebox(0,0){\vector(-1,0){1}}} 
  \put(  7,0.5){\makebox(0,0){2-form}} 
  \put(8.5,0.7){\makebox(0,0){curl}} 
  \put(8.5,0.4){\makebox(0,0){\vector(-1,0){1}}} 
  \put( 10,0.5){\makebox(0,0){1-form}} 
  \put(11.5,0.7){\makebox(0,0){grad}} 
  \put(11.5,0.4){\makebox(0,0){\vector(-1,0){1}}} 
  \put( 13,0.5){\makebox(0,0){0-form}} 
\end{picture}
  \hspace*{\fill}\\
In this scheme, the usual 6-point stencil of a discrete Laplacian operated
on a 0-form can be realized by div$\star$grad where the star denotes the Hodge
transform and the result is a dual 3-form. Likewise, the Laplacian on a
1-form can be written as div$\star$grad $-$ curl$\star$curl and returns a dual
2-form. Except for boundary effects, the primary and dual grids are
on equal footing.

For our problem, the magnetic field $\vec{B}$ is considered a
primary 1-form (or dual 2-form). The current density $\vec{j}$ then
is a primary 2-form. With these prescriptions, we can perform all
differential operation on scalar and on vector fields in a consistent
way.

We just mention in passing that these forms can be equally well
constructed on an irregular grid and element functions can be higher
order polynomials if more element parameters are adequately provided.
Hence high order difference formulas can be set up in a systematic way.
Multigrid extensions of the Whitney forms are a matter of current
research \citep{Gradinaru:2002}.

\section{The Wheatland-Sturrock-Roumeliotis (WSR) algorithm}
\label{sec:Wheatland}

The algorithm proposed by Wheatland, Sturrock, and Roumeliotis
(\citeyear{Wheatland:etal:2000}) tries to find
the force-free field $\vec{B}$ from the argument which minimizes
a penalty function $L$, i.e.,
\begin{equation}
  \vec{B} = \mathrm{argmin}(L)\;,\quad
  L(\vec{B}) = \int\limits_V |w\vec{j}\crss\vec{B}|^2
              + \int\limits_V |\div\vec{B}|^2\,.
\label{L_B}\end{equation}

In fact, the integrals can be looked upon as a Hilbert product on the
respective finite element space defined in the previous section.
This conception helps greatly when programming $L$ and its derivatives.
The calculation of $\div\vec{B}$ and $\crl\vec{B}$ throughout $V$ requires the
knowledge of $\vec{n}\sdot\vec{B}$ and $\vec{n}\crss\vec{B}$ on
the whole of the surface $\partial V$. This is more than (\ref{BC})
prescribes and the problem of minimising $L$ under these restrictive conditions
is clearly overdetermined. If inconsistent boundary conditions are
imposed on (\ref{L_B}), a decent minimum may never be reached.
We therefore allow for the option in our program to vary the normal
and/or tangential field components on individual faces of $V$.
This essentially is equivalent to setting $\vec{n}\sdot\vec{B}$
from $\div{\vec{B}}$ = 0 and $\vec{n}\crss\vec{B}$ consistent with a
vanishing Lorentz force on the respective boundary face.

In this sense the second integral in (\ref{L_B}) is calculated from
the squared (dual) 3-form $\div{\vec{B}}$ defined at the cell vertices by
summing over all vertices. Vertices on a domain surface, edge or
corner are especially weighted with factor 0.5, 0.25 and 0.125
respectively.

\begin{figure}
  \hspace*{\fill}
\begin{picture}(7.2,3.3)
  \put( 0.,3.3){\includegraphics[bb=70 53 220 770,clip,width=3cm,angle=-90]
            {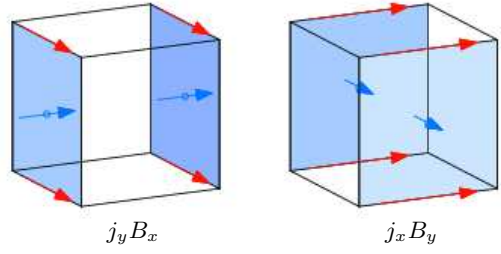}}
  \put(1.8,0.1){\makebox(0,0){\small $j_y B_x$}} 
  \put(5.5,0.1){\makebox(0,0){\small $j_x B_y$}} 
\end{picture}
  \hspace*{\fill}
\caption{Finite element parameters entering into the calculation of the
  $z$ component of $\vec{j}\crss\vec{B}$ at the cell centre. The {\it red
  arrows} on the cell edges denote magnetic field components, the {\it blue
  face centered arrows} the current components involved.}
\label{fig:FEjxB}\end{figure}

The $\vec{j}\crss\vec{B}$ integral is a little more involved.
Formally the exterior product of a 1-form and a 2-form should give a
3-form. Hence the $\times$ product here has not the
property of an exterior product in differential geometry. Yet a
3-form, however one for each component of $\vec{j}\crss\vec{B}$
individually, gives the most compact stencil for this expression. In
Figure~\ref{fig:FEjxB} we show the element parameters which are needed
for the $z$ component of the Lorentz force at a cell centre.
E.g., the contribution $j_x B_y$ is obtained from an average over
the two $x$-faces of the cell. For each of these faces, $j_x B_y$ is
calculated by multiplying $j_x$ at its centre with the average of the
two $B_y$ values from the two $y$-edges of this face. The integral in
(\ref{L_B}) then is a simple sum over all three components of the
Lorentz force. For each component, the respective squared 3-form
elements are summed over all cell centres.

In order to obtain argmin $L(\vec{B})$, WSR proposed
a simple Landweber scheme by iteratively advancing
\begin{equation}
\vec{B}^{(i+1)} = \vec{B}^{(i)} + s\,\delta\vec{B}^{(i)} \;,\quad
\delta\vec{B}^{(i)} = - \frac{\partial L}{\partial\vec{B}}(\vec{B}^{(i)})\,,
\label{Landweber}\end{equation}
which guarantees that $L$ decreases at every step provided step size
$s$ is small enough. In this scheme, improvements of $\vec{B}$ are
strictly along the negative gradient direction and step sizes are
only guessed and reduced if necessary.

We have instead implemented an unpreconditioned conjugate gradient iteration
which at every iteration step performs an exact line search to the
minimum of $L'(s)$ = $L(\vec{B}+s\,\delta\vec{B})$ along the
respective search direction. Moreover, it selects an improved search
direction $\delta\vec{B}$ instead of the gradient as in (\ref{Landweber}).

Note that in contrast to existing implementations of this scheme by
\citeauthor{Wheatland:etal:2000} (\citeyear{Wheatland:etal:2000}) or
Wiegelmann (\citeyear{Wiegelmann:2004}) who programmed the
discretization of the analytical derivative (\ref{Landweber}), we
always calculate the numerically more consistent derivative of the
discretized function $L$.

Conjugate gradient solvers are optimal for linear problems for which
the objective function $L$ depends to second order on the components
of $\vec{B}$. Our problem, however, is nonlinear and (\ref{L_B}) is
of fourth order in $\vec{B}$ through the $\vec{j}\crss\vec{B}$ term.
We make use of our formulation of (\ref{L_B}) on the special grid
introduced in the previous section in order to calculate all five
polynomial coefficients of $L'(s)$ in one go. This enables us to
perform the exact line search at every step without much effort
by a single function call.
From the new minimum, the new search direction $\delta\vec{B}^{(i+1)}$ is
chosen so that it is a descent direction and also $H$-orthogonal to the
previous search direction $\delta\vec{B}^{(i)}$. $H$ here is the local
Hessian $\partial^2 L/\partial B_i\partial B_j$ at the line search
minimum. Likewise, we can also choose the Hestenes-Stiefel variant
which yields a new search direction which is $H$-orthogonal with respect to
some average Hessian $H$.

A parameter still to be determined in (\ref{L_B}) is $w$. Its choice
will be discussed later along with the presentation of the results. In
fact, it will turn out to be favourable to make it a space-dependent
$w(\vec{x})$. This is another difference with respect to the
implementations of \citeauthor{Wheatland:etal:2000}
(\citeyear{Wheatland:etal:2000}) and Wiegelmann(\citeyear{Wiegelmann:2004}),
who took $w$ a function of $\vec{B}$.

\section{The Grad-Rubin (GR) algorithm}
\label{sec:GradRubin}

Several variations of this algorithm exist. While the previous
approach to find a force-free field reduced the problem to a formal
optimization procedure, the approach by Grad and Rubin
(\citeyear{Grad:Rubin:1958}) is inspired by a quasi-physical relaxation: at
any time in the iteration the current $\vec{j}$ =
$\alpha^{(n)}\vec{B}^{(n)}$ produces via the Biot-Savart law a new field
$\vec{B}^{(n+1)}$. According to the differences between
$\vec{B}^{(n+1)}$ and $\vec{B}^{(n)}$, $\alpha$ is then
redistributed along the new field lines giving rise to a new current
and hence a new Biot-Savart field.

In our code, we distribute $\alpha$ along given field lines by solving
\begin{equation}
   \vec{B}^{(n)}\sdot\grd{\alpha}^{(n)} = 0
\label{BgrdA}\end{equation}
for given $\vec{B}^{(n)}$ and boundary values for $\alpha$ on $\partial V$.
In the next step we correct the field by solving for the vector potential
$\delta\vec{A}$ of the field update
\begin{gather}
 \Delta\,\delta\vec{A} = \delta\vec{j}\,,
\label{D_A}\\
 \mathtext{where}\div{\,\delta\vec{A}}=0 \mathtext{and}
 \delta\vec{j} = \crl{\vec{B}^{(n)}} - \alpha^{(n)}\vec{B}^{(n)}\,,
\nonumber\end{gather}
with boundary conditions $\vec{n}\crss\,\delta\vec{A}$ = 0 and
$\ngrd{\,(\vec{n}\sdot\delta\vec{A})}$ = 0. These boundary conditions
insure that $\div{\,\delta\vec{A}}$ vanishes also on the domain boundaries.
In the subsequent field correction
\begin{equation}
 \vec{B}^{(n+1)} = \vec{B}^{(n)} + \crl{\,\delta\vec{A}}\,,
\end{equation}
the normal components of $\vec{B}$ on the domain boundaries remain
unchanged.

In terms of the forms introduced above, the scheme looks as follows:\\[5mm]
  \hspace*{\fill}
\begin{picture}(14.5,3)
  \put( 1.6,2.3){\makebox(0,0){primary grid:}}
  \put( 2,0.5){\makebox(0,0){dual grid:}}
  
  \put(  4,2.8){\makebox(0,0){$\phi$}}
  \put(  4,2.3){\makebox(0,0){0-form}}
  
  \put(5.5,2.5){\makebox(0,0){grad}}
  \put(5.5,2.4){\makebox(0,0){\vector(1,0){1}}}
  
  \put(  7,2.8){\makebox(0,0){$\vec{B}_{\mathrm{pot}}$, $\vec{B}$}}
  \put(  7,2.3){\makebox(0,0){1-form}}
  
  \put( 8.7,2.5){\makebox(0,0){curl -- $\alpha\,\sdot$}} 
  \put( 8.7,2.4){\makebox(0,0){\vector(1,0){1}}}
  
  \put(10.4,2.8){\makebox(0,0){$\delta\vec{j}$}} 
  \put(10.4,2.3){\makebox(0,0){2-form}}
  
  \put(11.9,2.5){\makebox(0,0){div}}
  \put(11.9,2.4){\makebox(0,0){\vector(1,0){1}}}
  
  \put(13.4,2.8){\makebox(0,0){0}}
  \put(13.4,2.3){\makebox(0,0){3-form}}
  
  \put(   7,1.4){\makebox(0,0){\vector(0,1){1.2}}}
  \put( 7.3,1.4){\makebox(0,0){+}} 
  
  \put(10.4,1.4){\makebox(0,0){\vector(0,-1){1.2}}}
  \put(10.9,1.4){\makebox(0,0){$\Delta^{-1}$}} 

  \put(  7,0.5){\makebox(0,0){2-form}} 
  \put(  7,0.0){\makebox(0,0){$\delta\vec{B}_{\mathrm{ind}}$}}
  
  \put(8.7,0.7){\makebox(0,0){curl}}
  \put(8.7,0.4){\makebox(0,0){\vector(-1,0){1}}}
  
  \put(10.4,0.5){\makebox(0,0){1-form}} 
  \put(10.4,0.0){\makebox(0,0){$\delta\vec{A}$}} 
\end{picture}
  \hspace*{\fill}\\[5mm]
The scheme basically starts form a potential solution 
on the left and then cycles the square at the centre. Any time
$\vec{B}$ has been updated, $\alpha$ is remapped using (\ref{BgrdA})
before the residual current $\delta\vec{j}$ is calculated. Due to
numerical discretization errors in the integration of $\alpha$ in
(\ref{BgrdA}), $\delta\vec{j}$ may have a spurious divergence which
is checked and if necessary, eliminated by iterating
$\div\delta\vec{j}$ = 0
a few times while preserving normal boundary conditions and
$\crl\delta\vec{j}$.

While the Poisson equation (\ref{D_A}) can efficiently be solved with
a multigrid solver, the major computational effort is spent in solving
(\ref{BgrdA}) to the required precision.
Since (\ref{BgrdA}) is first order, boundary values need only be
supplied at one end of the field line (or characteristic). With
(\ref{BC}) this is automatically satisfied, however this choice of
boundary conditions deprives us of the freedom to emphasize those
boundary areas where we assume the observations to be more reliable. We
therefore do not make the distinction between $(\partial V)^+$ and
$(\partial V)^-$ as in (\ref{BC}) but we safeguard our solver of
(\ref{BgrdA}) against inconsistent boundary values for $\alpha$ by
attaching a weight with every boundary value for $\alpha$. The final
value on the characteristic is the according weighted average from
both end points.
This way, the influence from uncertain boundary values on the side
walls or from imprecise measurements on the bottom (photospheric)
boundary can be suppressed.

In principle, (\ref{BgrdA}) is solved by mapping every cell centre
along a field line to the boundary and interpolating the boundary
values to its foot point. To minimise the number of field line
calculations, we store the $\alpha$ value also in every cell the
calculated field line intersects along with the intersection
coordinates. For about 2/3 of the cells, the field line calculation
from their centre then can be discarded, because they have previously
been intersected by so many field lines, that their $\alpha$ value can
reliably be interpolated form the intersection information stored.

\section{Results}
\label{sec:Results}

We have tested our codes with two different model fields. The first is
the Low and Lou (\citeyear{Low:Lou:1990}) field model, one of the few analytic
force free field solutions which now has almost become a standard for
tests of force-free reconstruction schemes.
For the second test model, a twisted flux tube, we do not have an
analytic solution. For the existence  of a solution we rely on the
symmetry of the boundary conditions supplied.

\subsection{Low and Lou model}
\label{results:LowLou}

\begin{figure}[h!]
  \hspace*{\fill}
\includegraphics[bb=80 180 530 600,clip,width=7.5cm,angle=-90]{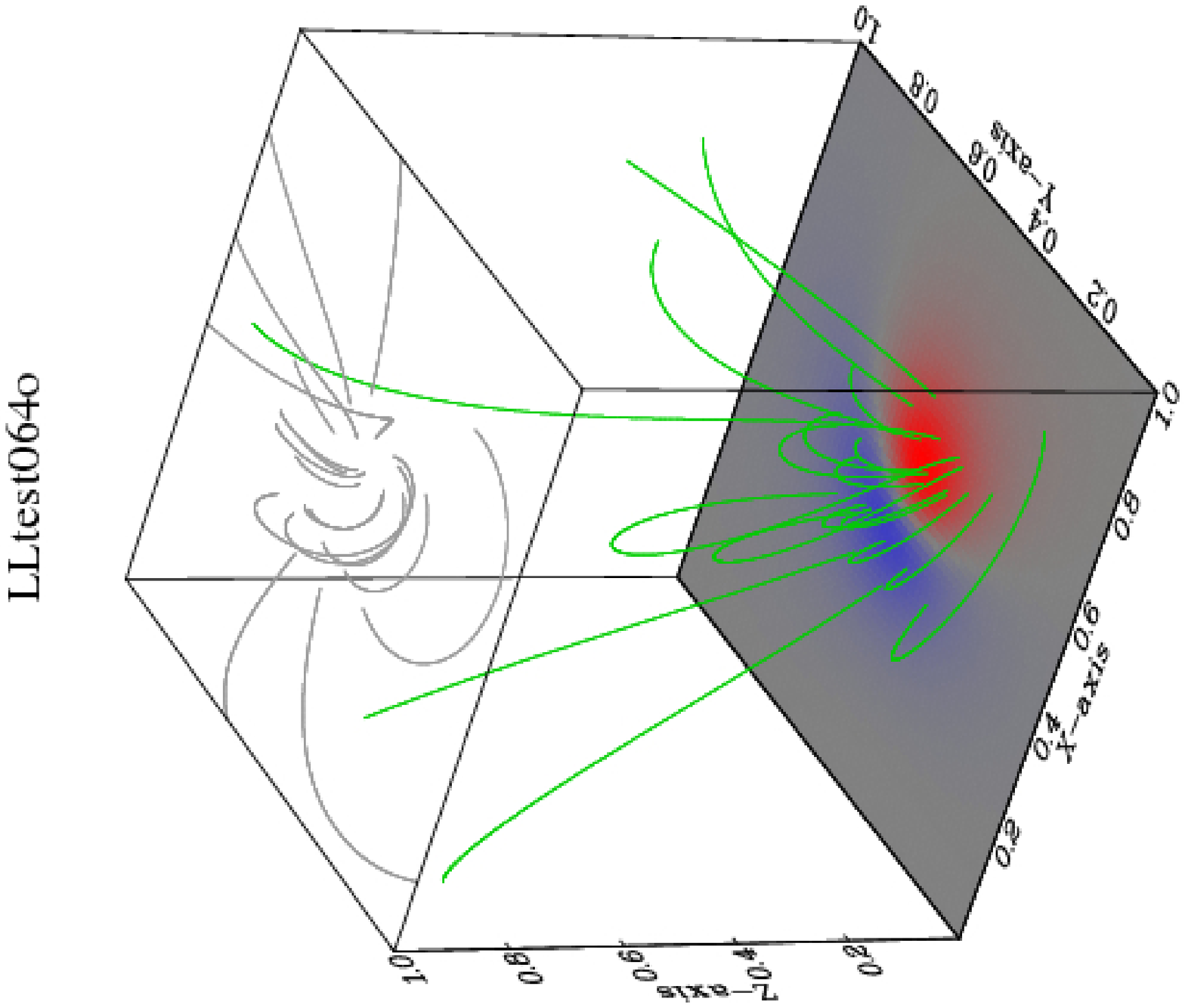}
\hspace{2mm}
\includegraphics[bb=80 180 530 600,clip,width=7.5cm,angle=-90]{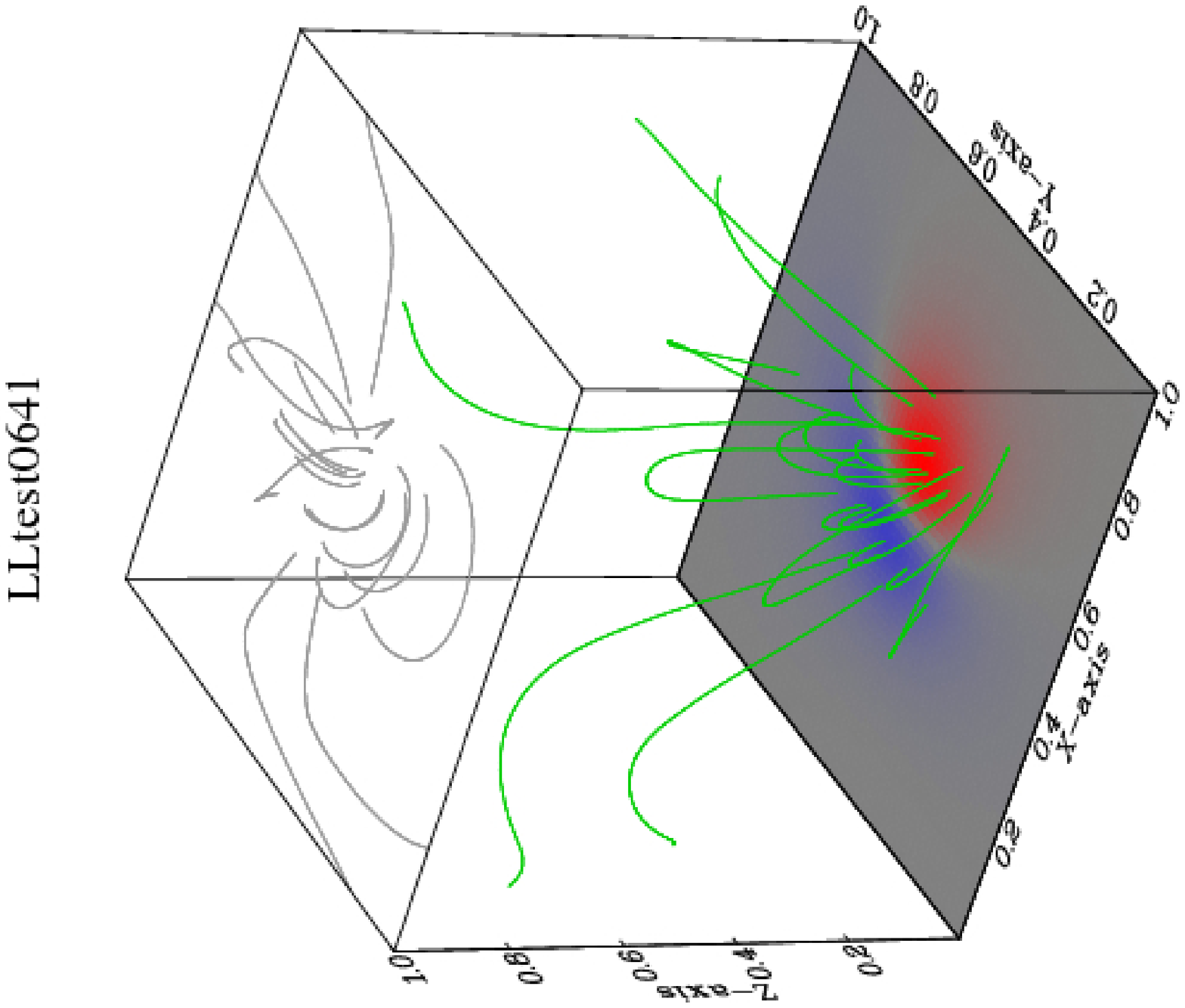}
  \hspace*{\fill}\\
  \hspace*{\fill}
\includegraphics[bb=80 180 530 600,clip,width=7.5cm,angle=-90]{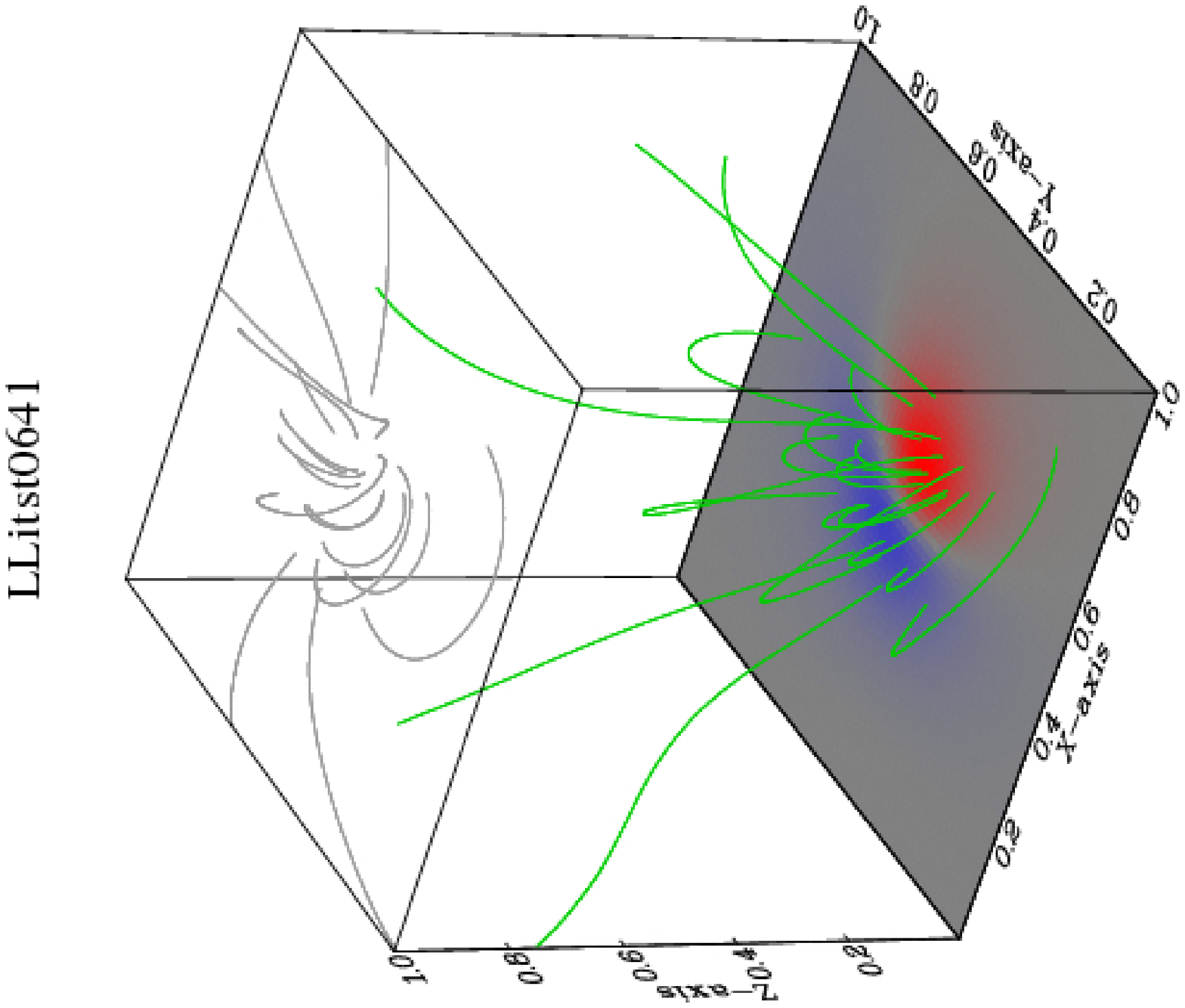}
\hspace{2mm}
\includegraphics[bb=80 180 530 600,clip,width=7.5cm,angle=-90]{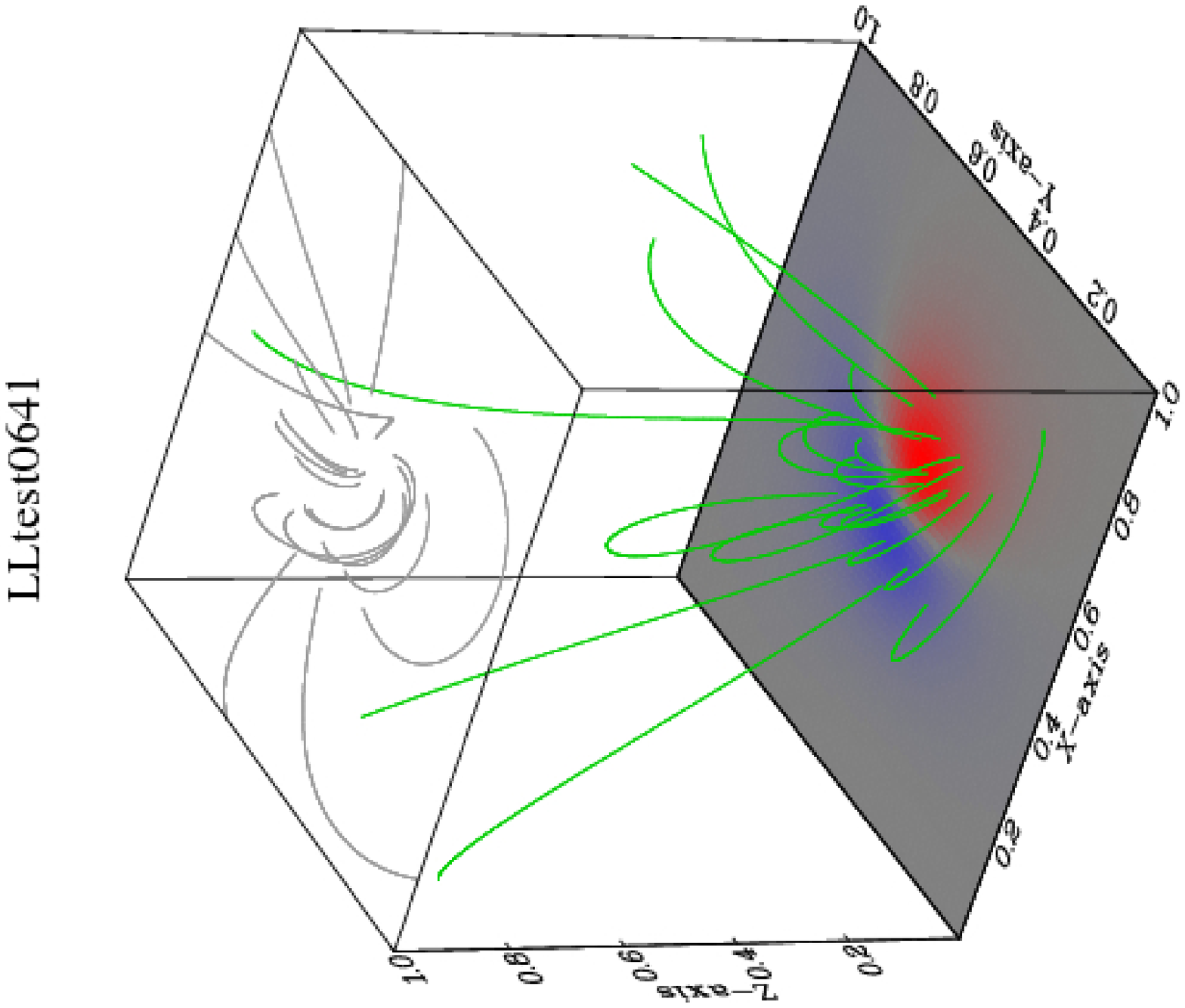}
  \hspace*{\fill}\\
  \caption{ Representative field lines of the original Low and Lou (1990)
    model ({\it upper left}), the result of the WSR iteration with
    $w$ = const({\it upper right}), the result of the WSR iteration with
    $w$ = $1/|B_{\mathrm{pot}}|$ ({\it lower left}), the result
    of the GR iteration ({\it lower right}). The {\it colour code} at the
    bottom represents $B_z$, the {\it top plane} shows the vertical
    projection of the field lines.}
\label{fig:PltFln1}\end{figure}

As a test case model we use the analytical nonlinear force-free field
solution of Low and Lou (\citeyear{Low:Lou:1990}). The field results form a
multipole with eigenvalue $a^2$ = 0.42659. We placed it at a depth of
$l$ = 0.2 below the centre of the ground plane and with orientation in
the $x$-$z$ plane with 45 degrees inclination to the $x$ and $z$ axes.
The field was scaled so that at the ground plane the vertical field
strength ranges between -9.95 and 5.09, the alpha values between
-16.9 and 5.61 and the vertical current density between -67.9 and 106.
A field line plot of this model is shown in the upper left of
Figure~\ref{fig:PltFln1}.

The result of the extrapolation is displayed in the other three panels,
all for a grid size $n$ = 64.
We find that the WSR iteration is strongly biased 
towards regions where the field strength is large. This very probably is
due to the Lorentz force term in (\ref{L_B}) which increases with
${\cal O}(B^4)$. For this reason, the original object function $L$
proposed by WSR \citep{Wheatland:etal:2000} had $w$ $\sim$ $1/|\vec{B}|$.
This choice, however, makes $L$ a rational function of the $B$ components
which we expect to slow down the convergence of the minimization
iteration. We therefore prefer the choice $w$ $\sim$
$1/|\vec{B}_{\mathrm{pot}}|$ where $\vec{B}_{\mathrm{pot}}$ is the
potential field consistent with the normal component of the given
boundary magnetic field. The potential field is obtained easily and
our choice of $w$ similarly reduces the influence of regions where the
field is expected to be strong. Since $w$ is not changed during the
iteration, it does not destroy the analytic properties of
$L(\vec{B})$.

The effect of this weighting is remarkable, since $B_{\mathrm{pot}}$
varies by three orders of magnitude between 0.027 and 28 throughout $V$.
In Figure~\ref{fig:PltFln1} we show the result of the WSR
iteration with $w$ = const (upper right) and $w$ =
const$/|\vec{B}_{\mathrm{pot}}|$ (lower left). The discrepancy of
the former at larger heights in regions of small magnetic field strength
is immediately apparent.
The lower right panel shows the result of the GR iteration. It
reproduces the original in weak and strong field regions equally well
to at least the plotting precision. Details in the original field lines
are reproduced exactly.

\begin{figure}[t]
  \hspace*{\fill}
\includegraphics[bb=120 50 464 745,clip,height=12cm,angle=-90]{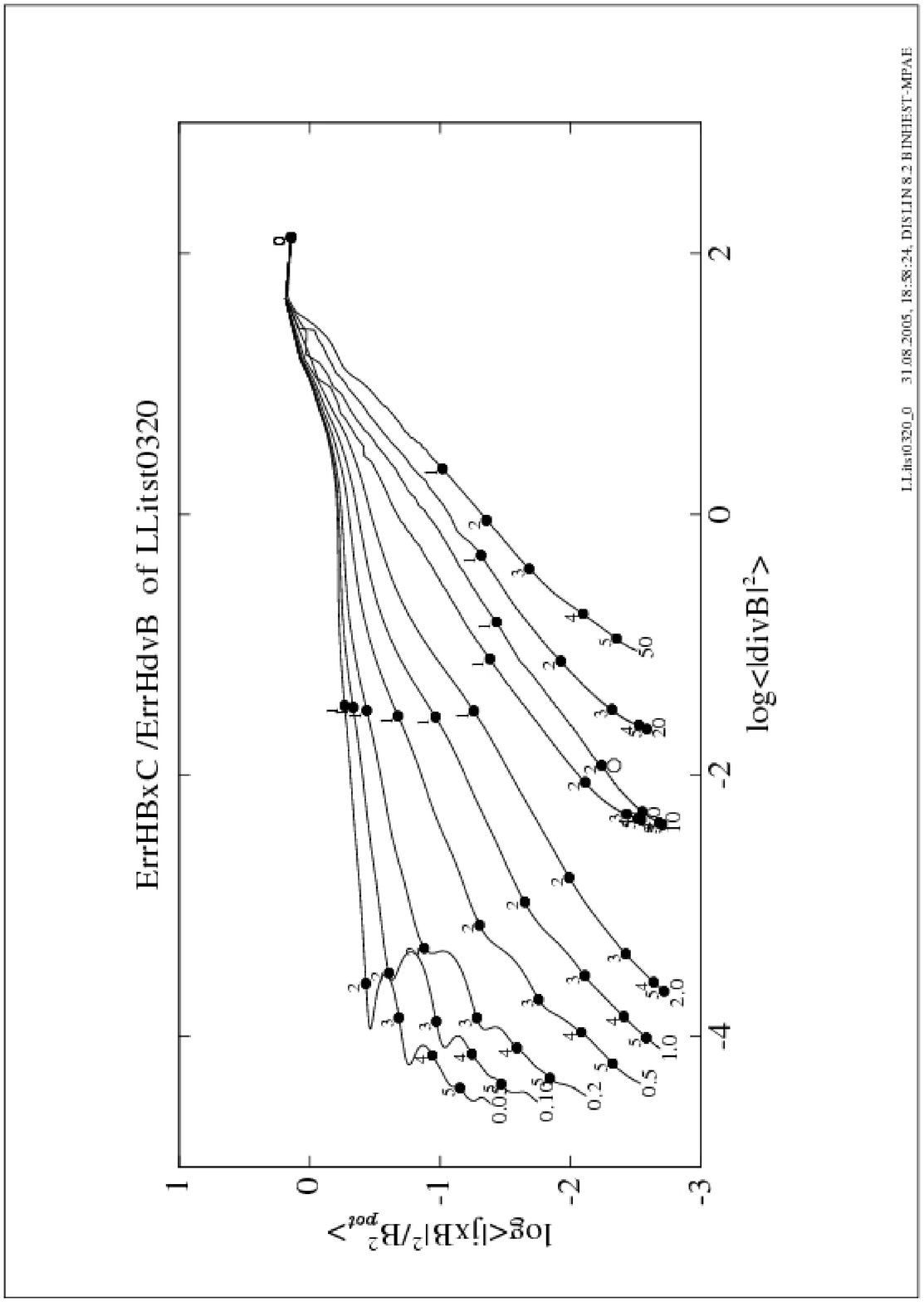}
  \hspace*{\fill}\\
  \hspace*{\fill}
\includegraphics[bb=80 50 530 745,clip,height=12cm,angle=-90]{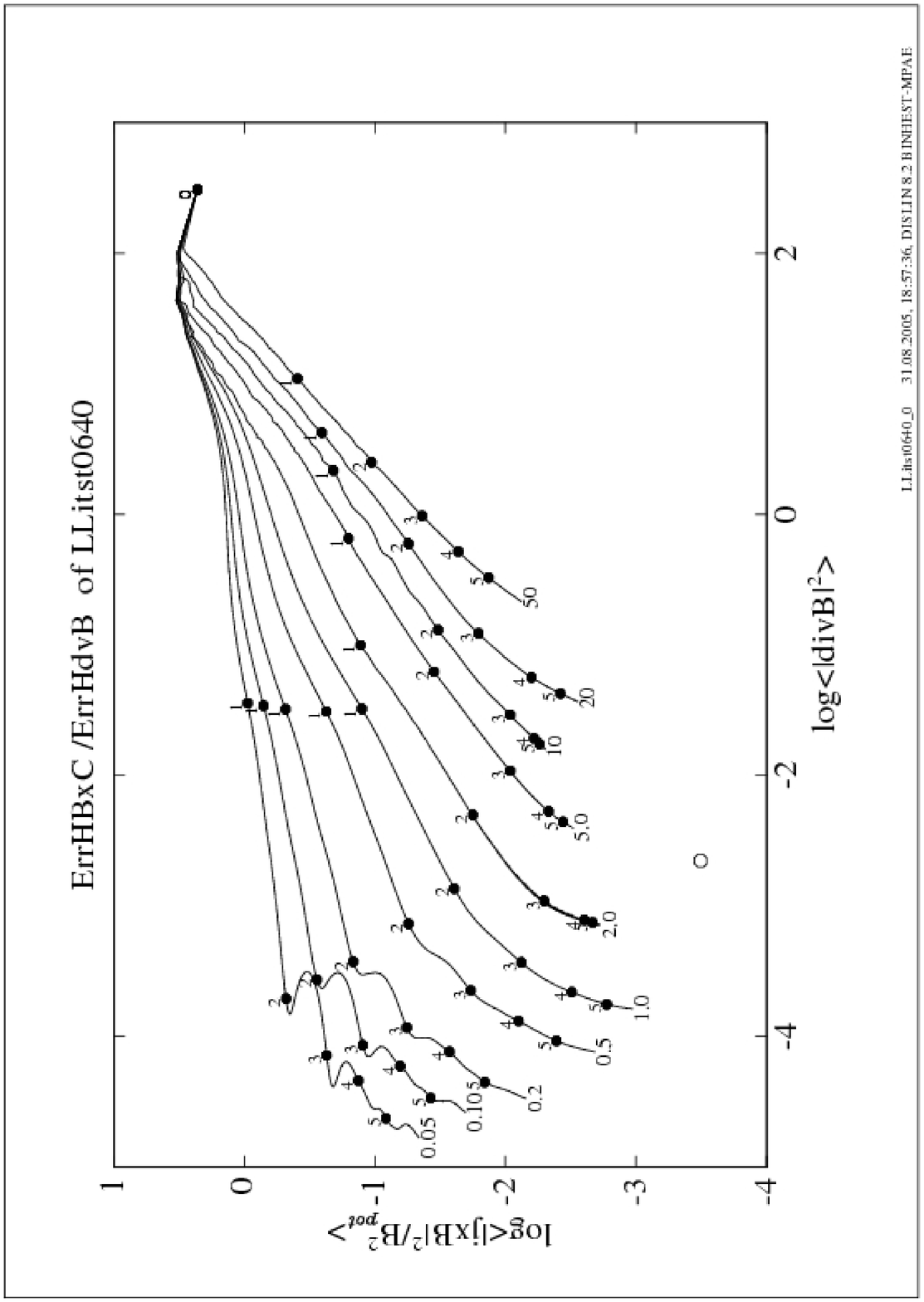}
  \hspace*{\fill}\\
  \caption{Decrease of the two integral terms in (\ref{L_B}) during the
    conjugate gradient iteration for a grid $n$ = 32 ({\it top}) and
    $n$ = 64 ({\it bottom}).
    The weight function was chosen $w$ = const/$|\vec{B}_{\mathrm{pot}}|$
    with different values for the constant as marked at the {\it lower left
    end} of the curves. The {\it tick marks} along the curve indicate every
    50${th}$ ($n$=32) or 100${th}$ ($n$=64) iteration step.
    The {\it circle} in each diagram indicates the integral values
    of the analytic model calculated on the respective grid.}
\label{fig:PltHst}\end{figure}

For the WSR iteration the question remains how to choose
the constant in the weighting factor $w$. In Figure~\ref{fig:PltHst} we display
the iteration history of the two integral terms in (\ref{L_B}) for different
constants in $w$ = const$/|\vec{B}_{\mathrm{pot}}|$ and for
two different grid sizes $n$ = 32 and 64.
The iterations were started with an initial $\vec{B}^{(0)}$ = 0 inside $V$
and the right boundary conditions for the normal and tangential fields
on $\partial V$.
Note that for $\vec{B}^{(0)}$ the divergence and the Lorentz force do not
vanish because of these boundary values imposed.

A value const $\sim$ 1 seems optimal which is not surprising because
with this choice of $w$ both terms are of the same order $B^2/h$.
Smaller values of the constant lead to a faster reduction of the
div$\vec{B}$ term, larger values yield a bias towards the elimination
of the Lorentz force. In Figure~\ref{fig:PltHst} we have chosen the axes
scales for the ordinate and the abscissa equally, so that the curve
(const $\sim$ 20) which decreases closest to 45 degrees eliminates
both integral terms at about equal rates.

For the case $w$ = const similar plots can be produced as in
Figure~\ref{fig:PltHst} and the optimal constant then is around 0.05. It
depends, however, on the magnetic field strength of the model since
now the two terms in (\ref{L_B}) have different units.

The number of iterations performed in Figure~\ref{fig:PltHst} was
10$\times n$ where $n$ is the number of cells along each coordinate axis.
From the similarity of both diagrams we conclude that roughly the number
of iterations needed to reduce $L$ by a given factor increases proportional
to $n$. This result seems to differ from previous
implementations of the WSR algorithm which used a
Landweber iteration with a more or less effective step size control
\citep{Schrijver:etal:2006}.
For these codes, the number of iterations was found to increase with
$n^2$. Note, however, that these authors measured the number of steps
until the magnitude of $\delta\vec{B}$ fell below a given limit
rather than the convergence speed (i.e., the fractional decrease of $L$ per
iteration step).

Of course, the precision of the analytic model, calculated on the
respective grid positions increases with higher grid resolution (see
circles in the diagrams in Figure~\ref{fig:PltHst}). For the $n$ = 32 grid
this precision is reached already after 100 iterations with const = 10,
while for $n$ = 64 the much higher analytical precision seems to be
out of reach.

\begin{figure}
  \hspace*{\fill}
\includegraphics[bb=55 100 557 700,clip,height=12cm,angle=-90]{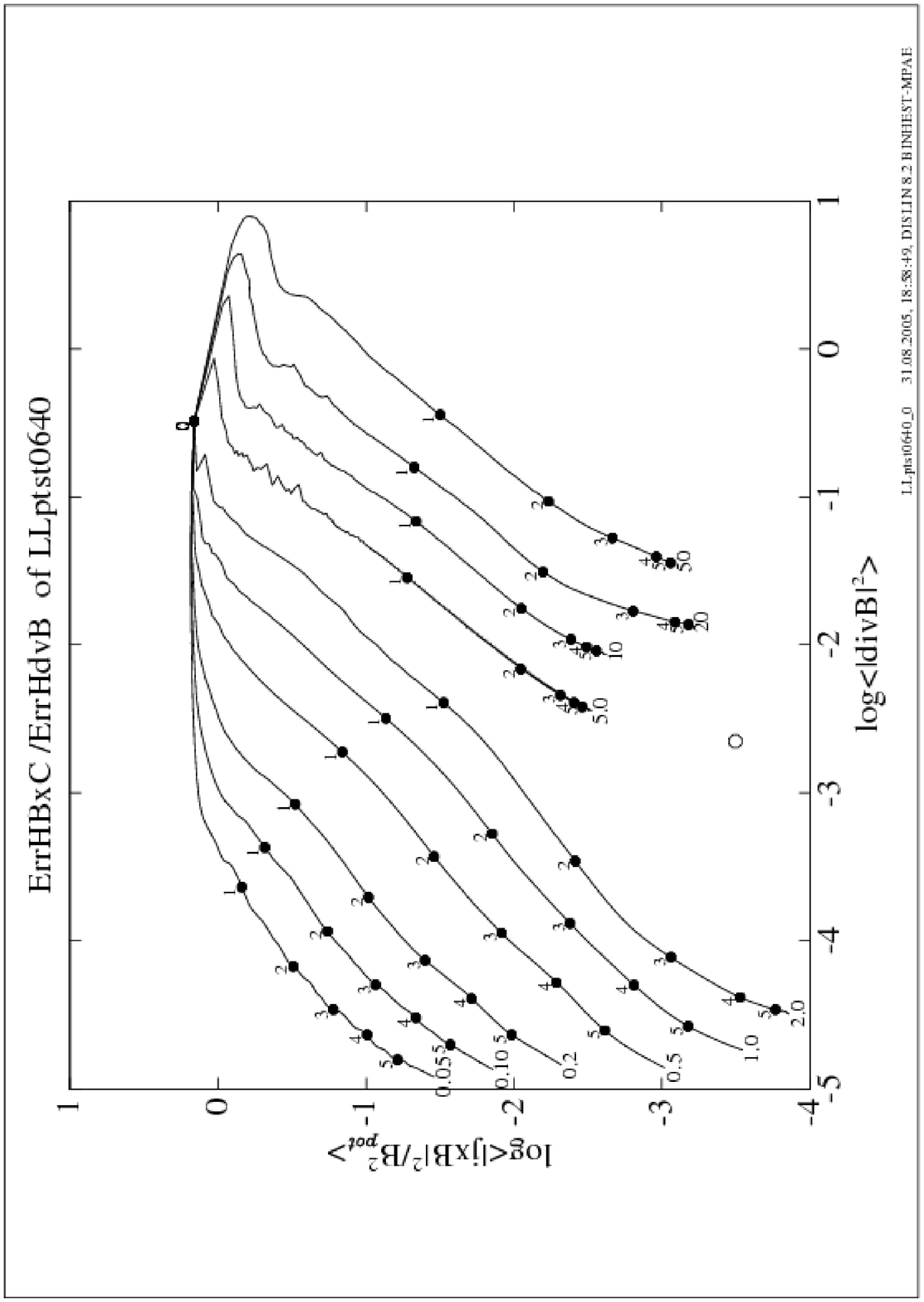}
  \hspace*{\fill}\\
  \caption{Same as lower diagram in Figure~\ref{fig:PltHst}, but with
    a potential field as initial $\vec{B}$ for the iteration.}
\label{fig:PltHstp}\end{figure}

It is well known that it depends among other details also on the
initial iterate $\vec{B}^{(0)}$ how closely a desired solution can be
approached by Krylov-type iteration schemes (e.g., Landweber, steepest
descent or conjugate gradients).
During the iteration, the search directions $\delta\vec{B}^{(i)}$
build up a subspace of the Hilbert space for $\vec{B}$ and
discrepancies between the iterate $\vec{B}^{(i)}$ and the solution
can at best be eliminated inside this subspace, usually called Krylov
space \citep{Saad:2003}. Differences between $\vec{B}^{(0)}$ and the
solution which fall out of the Krylov space cannot be corrected.
A good choice of the initial $\vec{B}^{(0)}$ therefore does not only
save computation time but may be a necessity to reach a solution at all, 
at least for ill-posed problems for which the Krylov space remains limited.
In Figure~\ref{fig:PltHstp} we show a diagram similar to the $n$ = 64 case in
Figure~\ref{fig:PltHst}, but here the initial $\vec{B}^{(0)}$ was chosen to
be the potential field inside $V$ for the given normal component boundaries 
but with the tangential field components on $\partial V$ expected for the
final force-free field.
Note that again the Lorentz force and the divergence of the initial
$\vec{B}^{(0)}$ do not vanish because the final, non-potential tangential
field boundary values on $\partial V$ have been enforced.

The diagram shows that with this improved initial field, it is no
problem to reach the error bounds of the analytical model if const $<$ 5.
Note, however, a larger value for the constant, i.e., a greater bias towards
the Lorentz force in (\ref{L_B}) during the iteration does not
necessarily produce a smaller final error of the Lorentz force term. 

For the GR scheme, the iterates $\vec{B}^{(i)}$ are always
numerically divergence free and at any step they satisfy the normal
boundary conditions. We therefore cannot start this iteration from
$\vec{B}^{(0)}$ = 0 but take the potential field as a starting point
instead. Since the information about the tangential boundary field is
stored in the boundary values of $\alpha$, the divergence of
$\vec{B}^{(0)}$ and the current $\vec{j}^{(0)}$ exactly vanish in
contrast to the WSR initial field.

\begin{figure}
  \hspace*{\fill}
\includegraphics[bb=173 60 532 380,clip,width=7.5cm,angle=-90]{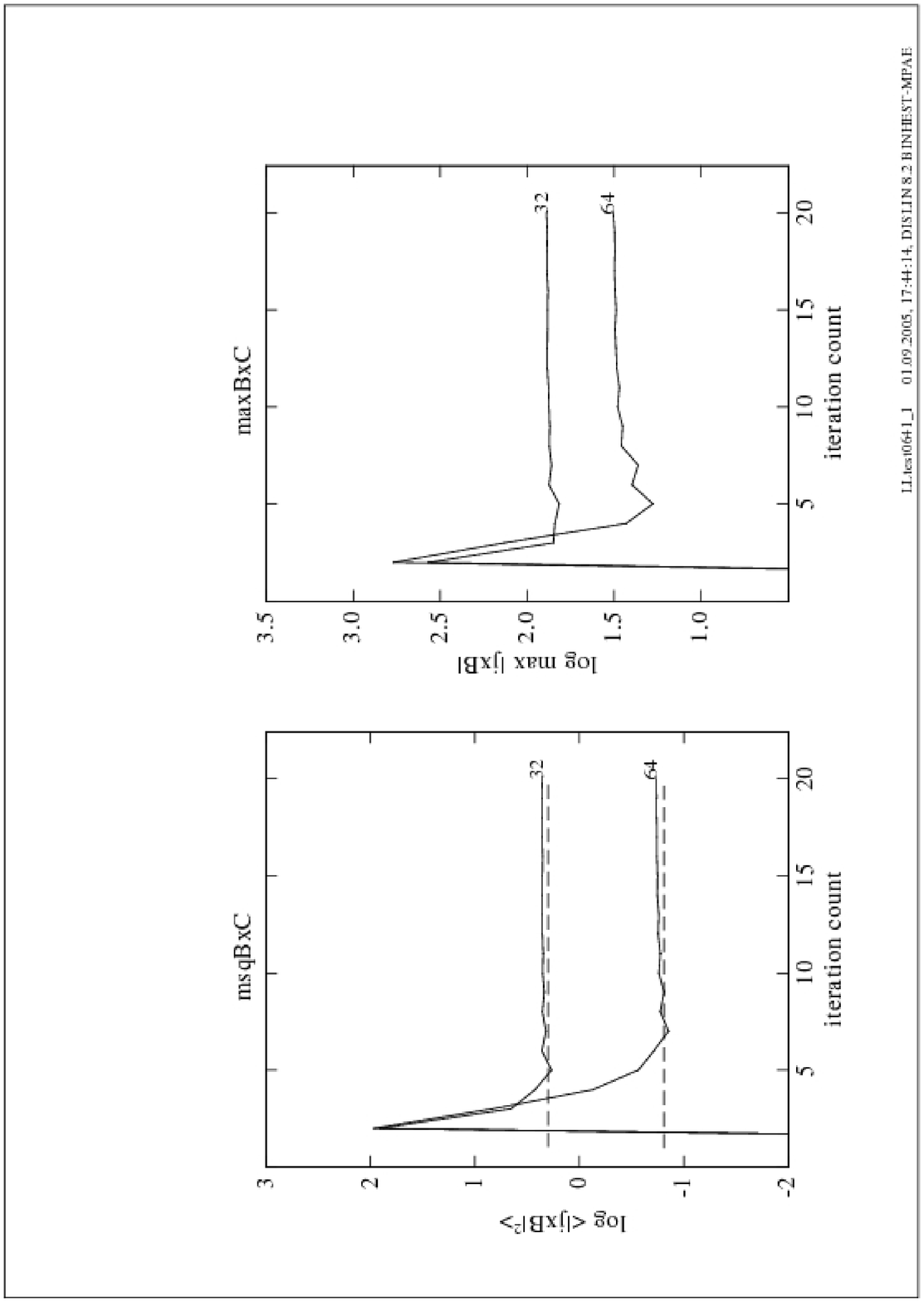}
\includegraphics[bb=173 84 532 430,clip,width=7.5cm,angle=-90]{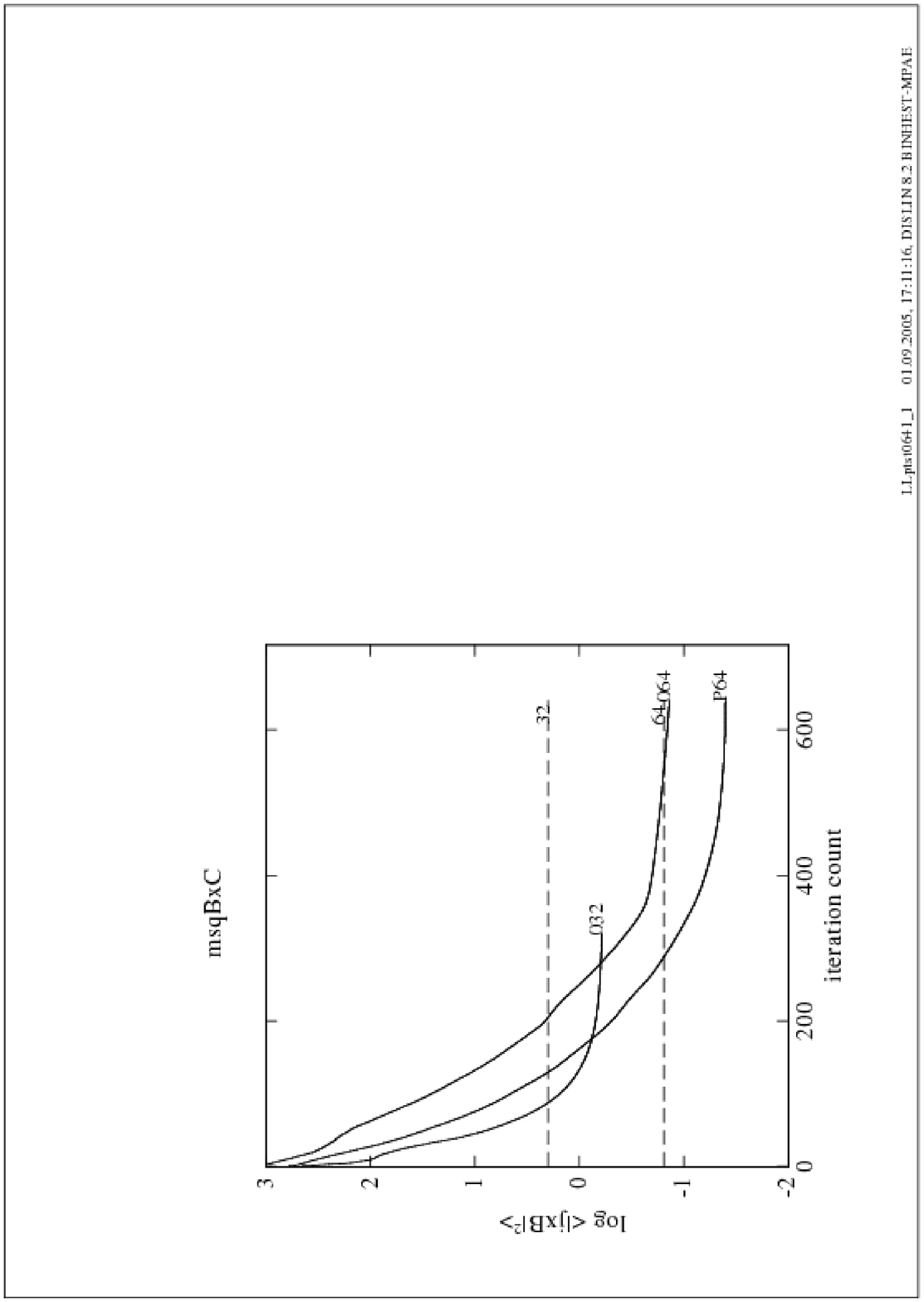}
  \hspace*{\fill}\\
  \caption{Mean square Lorentz force
    $\bra|\vec{j}^{(i)}\crss\vec{B}^{(i)}|^2\ket$ in the course
    of the GR ({\it left}) and the WSR ({\it right}) iteration.
    For the WSR iteration the constant was chosen to be 2.
    The {\it curves} are the result for the $n$ = 32 and 64
    grid as marked.
    For the latter we show results for an initial $\vec{B}^{(0)}$ = 0 (064)
    and $\vec{B}^{(0)}$ = $\vec{B}_{\mathrm{pot}}$ (P64) except for the
    boundary values.
    The {\it dashed lines} denote the respective residual Lorentz force
    for the discretized Low and Lou (1990) solution
    $\vec{B}_{\mathrm{org}}$.}
\label{fig:msqBxC}

  \hspace*{\fill}
\includegraphics[bb=173 400 532 720,clip,width=7.5cm,angle=-90]{GRBxC.eps}
\includegraphics[bb=173  83 532 430,clip,width=7.5cm,angle=-90]{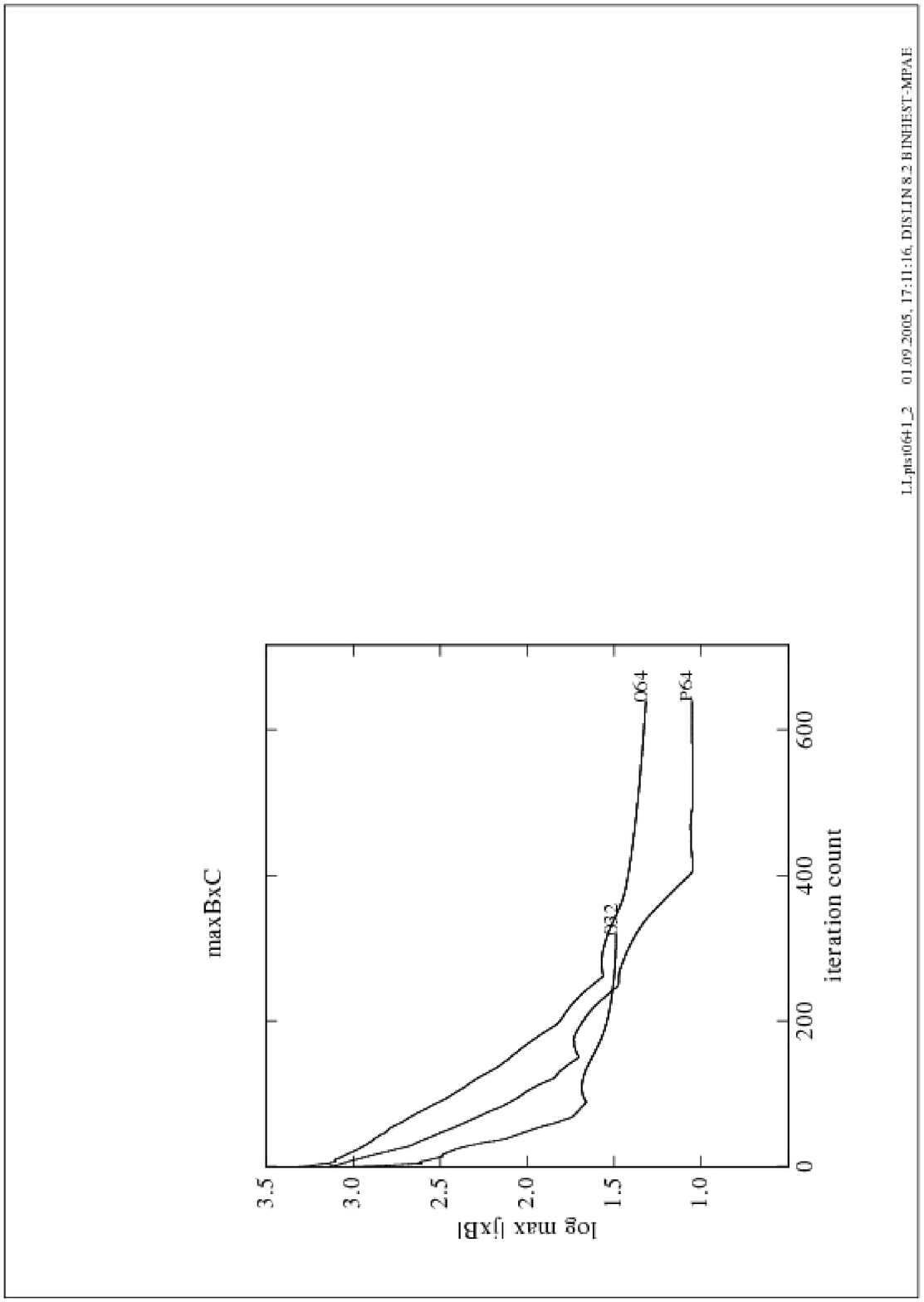}
  \hspace*{\fill}\\
  \caption{Same as Figure~\ref{fig:msqBxC}, but the maximum Lorentz force
    max$|\vec{j}^{(i)}\crss\vec{B}^{(i)}|$ in $V$
    for the GR iteration ({\it left}) and the WSR
    scheme ({\it right}).}
\label{fig:maxBxC}\end{figure}

The GR scheme converges much more rapidly than the
WSR algorithm. However, the GR iteration scheme does not
guarantee a continuous decrease of a certain object function, such
as $L$ for the WSR case. Less than 10 iterations were
needed to eliminate the residual Lorentz forces to the level of the
discrete Low and Lou solution (see Figure~\ref{fig:msqBxC} and
\ref{fig:maxBxC}). However, once this level has been reached, the Lorentz
forces could not be lowered any further but even increased slightly
again. This holds both for the mean and the maximum residual Lorentz
force.
Note that in Figs.~\ref{fig:msqBxC} we show the mean square Lorentz force
without weight $w$ $\sim$ $1/|B_{\mathrm{pot}}|$ as in Figure\ref{fig:PltHstp}.

From our calculations we may conclude therefore that with a WSR code
and a potential field as initial iterate smaller residual Lorentz
forces may eventually be reached than with the GR approach, provided
the WSR code is supplied with consistent and exact boundary conditions
of, e.g., a known analytic force free field. In this case a minimum
$L$ = 0 exists and we are confident that our code will approach
towards it due to its exact line search capabilities. The number of
iterations though may be prohibitive in some cases.

\begin{figure}
  \hspace*{\fill}
\includegraphics[bb=173 60 532 380,clip,width=7.5cm,angle=-90]{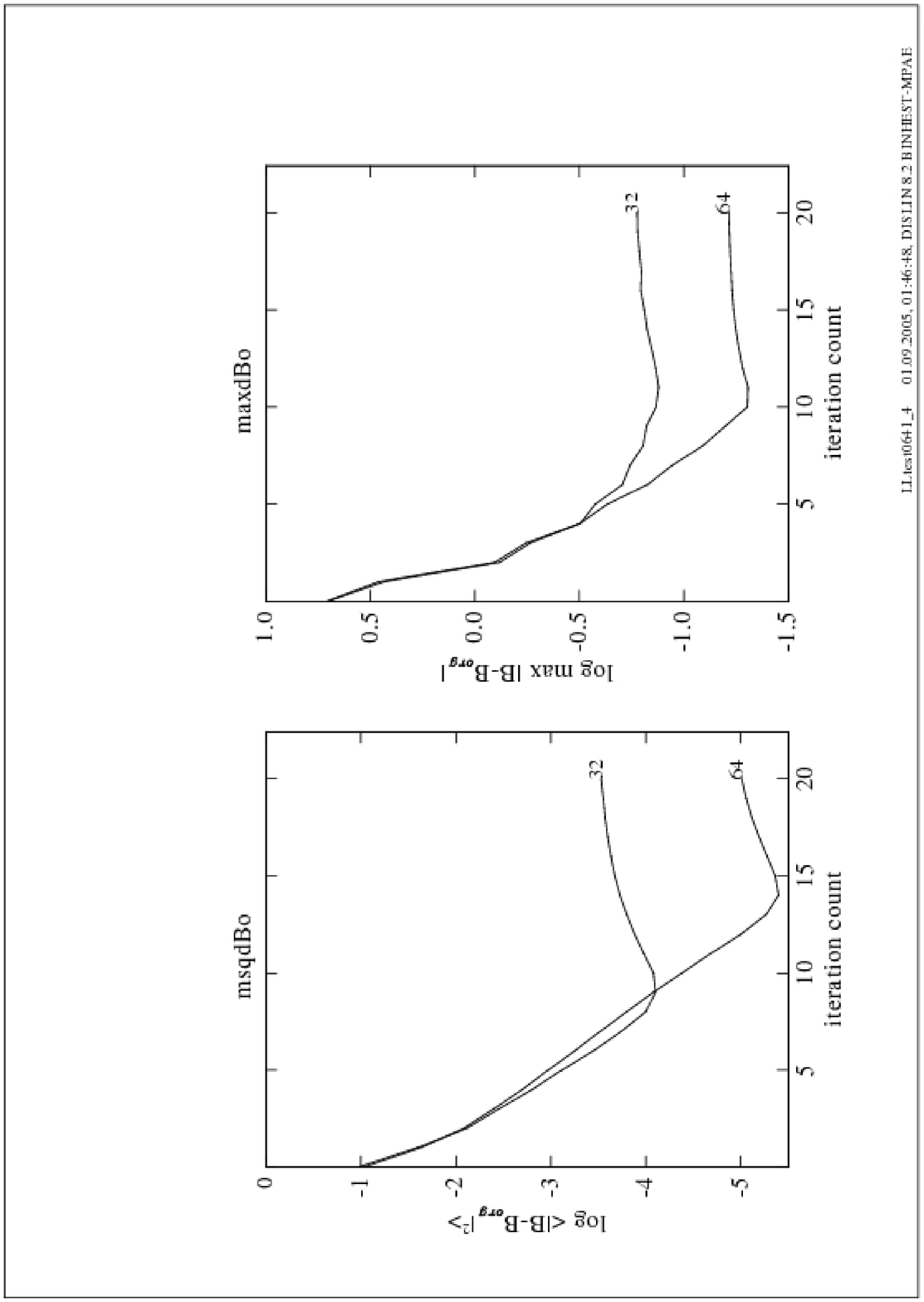}
\includegraphics[bb=173 84 532 430,clip,width=7.5cm,angle=-90]{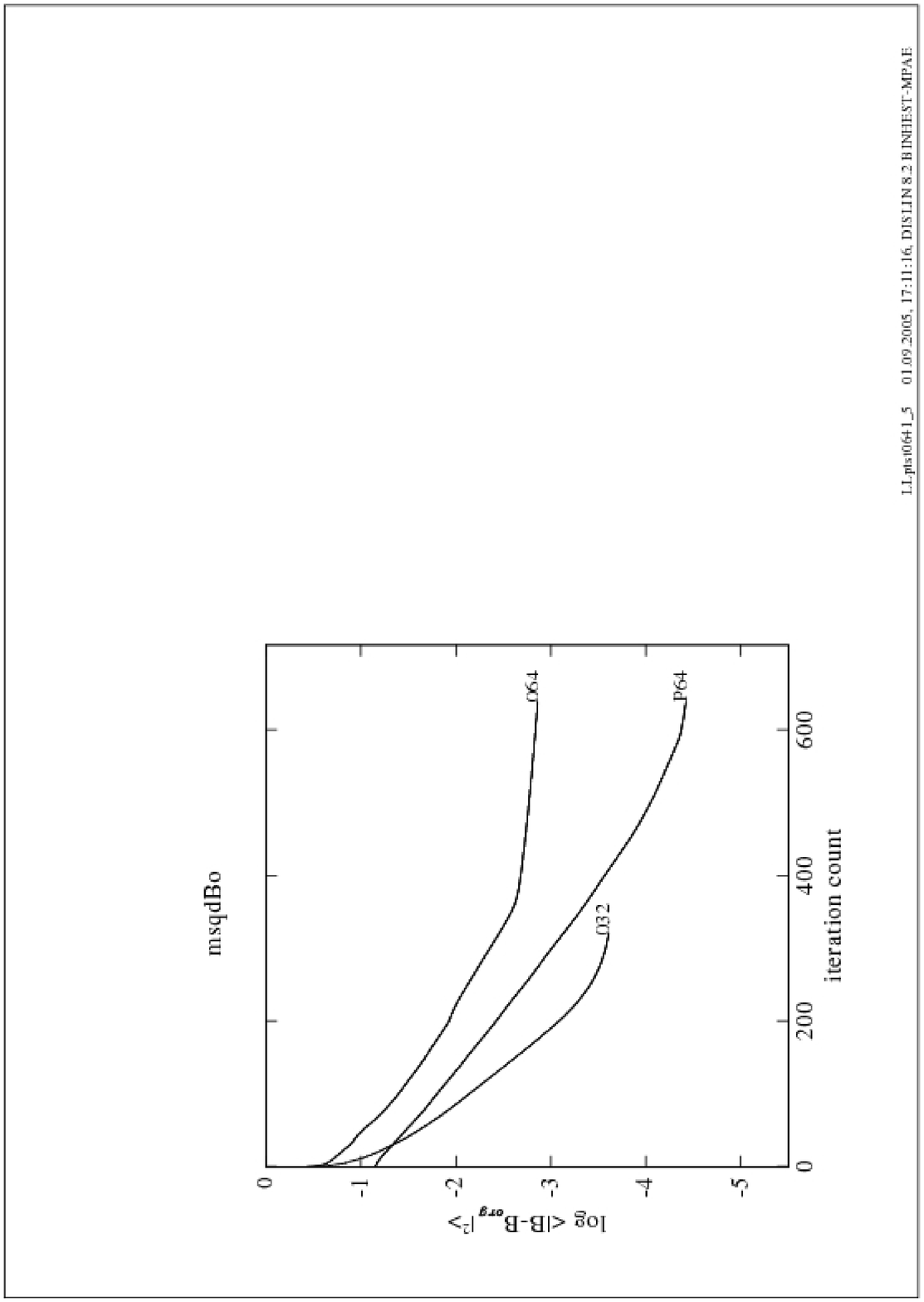}
  \hspace*{\fill}\\
  \caption{Mean square difference
    $\bra(\vec{B}^{(i)}-\vec{B}_{\mathrm{org}})^2\ket$
    of the iterated field to the original Low and Lou (1990) solution
    for the GR iteration ({\it left}) and the WSR
    scheme ({\it right}). The coding of the curves refer to the
    same calculations as in Figures~\ref{fig:msqBxC} and \ref{fig:maxBxC}.}
\label{fig:msqdBo}

  \hspace*{\fill}
\includegraphics[bb=174 400 532 720,clip,width=7.5cm,angle=-90]{GRdBo.eps}
\includegraphics[bb=174  79 532 430,clip,width=7.5cm,angle=-90]{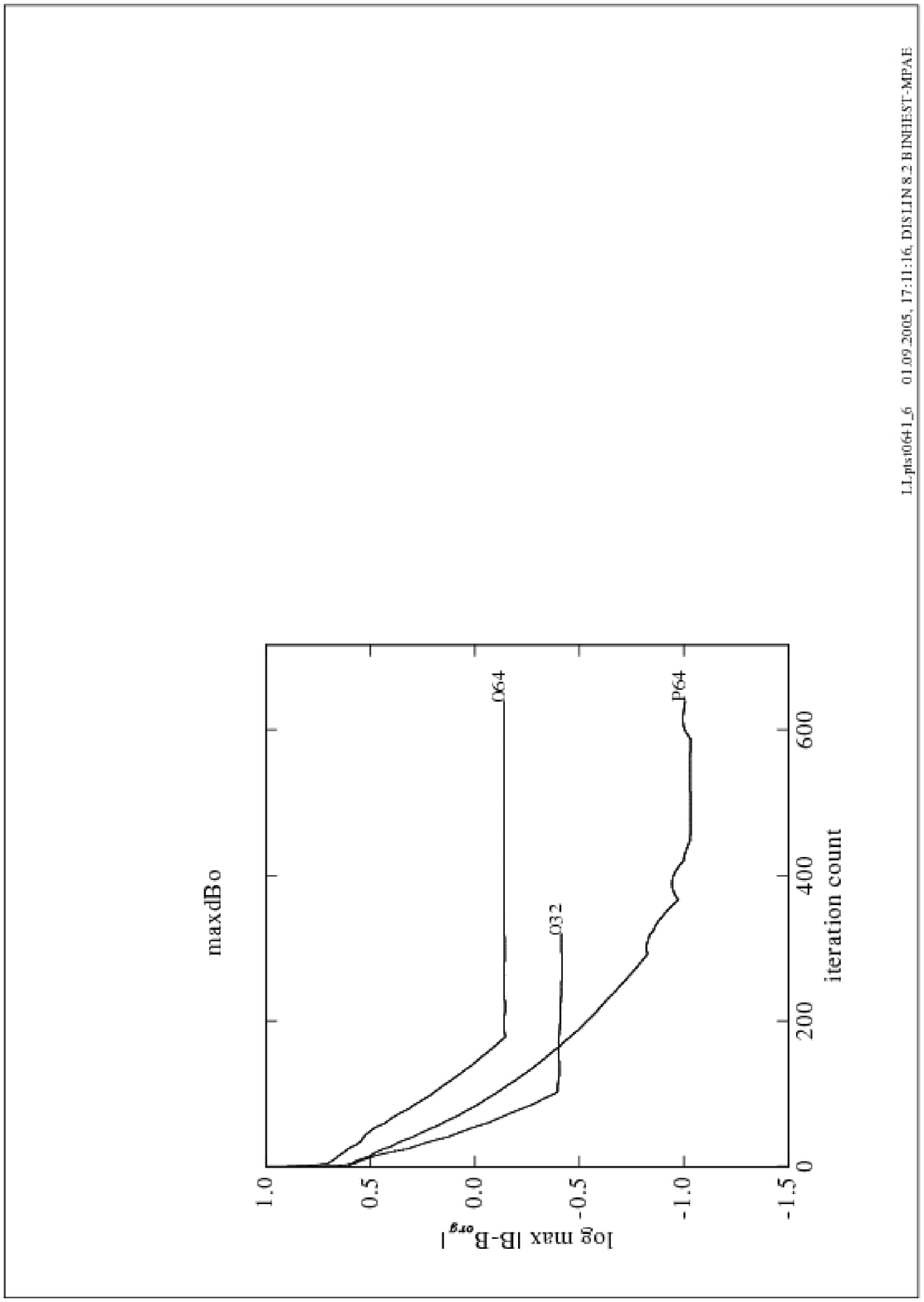}
  \hspace*{\fill}\\
  \caption{Same as Figure~\ref{fig:msqdBo}, but the maximum difference
    max$|\vec{B}^{(i)}-\vec{B}_{\mathrm{org}}|$
    of the iterated field to the original Low and Lou (1990) solution
    for the GR iteration ({\it left}) and the WSR
    scheme ({\it right}).}
\label{fig:maxdBo}\end{figure}

We have seen in the field line plots, that the GR result, even though
it retains residual Lorentz forces, approaches the discrete Low and
Lou solution extremely well. This can be confirmed if the difference
$\vec{B}^{(i)}-\vec{B}_{\mathrm{org}}$ is monitored during the
iteration.
As is shown in Figs.~\ref{fig:msqdBo} and \ref{fig:maxdBo}
the GR code here performs much better than WSR.
The mean square error it achieves is almost an order of magnitude smaller
than the WSR code can reach, even with the potential initial
field.
It seems that the most consistent field with the least Lorentz forces
is reached after only 5 ($n$ = 32) or $\sim$ 7 ($n$ = 64) iterations.
Continuation of the iteration brings the field still closer to the
Low and Lou solution but increases the level of the Lorentz force
slightly.
After about 10-14 iterations the difference $\vec{B}^{(i)} -
\vec{B}_{\mathrm{org}}$ is minimized and rises thereafter. A critical
point for a practical application of the GR code is therefore when
to stop the iteration. The saturation of the residual Lorentz
forces seems to be a helpful indication.

To measure differences of our final iterate with the original Low and
Lou field, we calculate the following two error norms:
\[
E_D=\frac{\sum\limits_{\mathrm{elements}}\;\sum\limits_{i\in\{x,y,z\}}
                                    |B_i-(\vec{B}_{\mathrm{org}})_i|
         }
         {\sum\limits_{\mathrm{elements}}\;\sum\limits_{i\in\{x,y,z\}}
                         |(\vec{B}_{\mathrm{org}})_i|
         }\,,
\]\[
E_C=\frac{\sum\limits_{\mathrm{elements}}\;\sum\limits_{i\in\{x,y,z\}}
                          |B_i\sdot(\vec{B}_{\mathrm{org}})_i|
         }
         {\sum\limits_{\mathrm{elements}}\;\sum\limits_{i\in\{x,y,z\}}
                              (\vec{B}_{\mathrm{org}})_i^2
         }\,.
\]
$E_D$ measures a normalized difference to the original field,
$E_C$ a correlation with it. The two error measures are not independent:
for small errors, $1-E_C$ $\sim$ $E_D^2$.
Note that we do not subtract or correlate vectors but every individual
component because from our grid, we obtain the values for different components
at different locations. 
For the final GR result on a $n$ = 64 grid, we obtain
\[
    E_D = 3.81\;10^{-3} \;;\quad
    E_C = 1.000034
\]
after 14 iteration steps which took 18 minutes of calculation on a 667 MHz
Pentium IV Linux operated single processor with 256 MB Ram.
The residual Lorentz forces of the reconstructed field had the same level
as the discretized original Low and Lou field and the divergence was of the
order of the numerical roundoff error.

The equivalent WSR computations starting from  $\vec{B}^{(0)}$ =
$\vec{B}_{\mathrm{pot}}$ yield
\[
   E_D = 1.81\;10^{-2} \;;\quad
   E_C = 0.99947
\]
after 640 iteration steps and 22 minutes of calculation time on the same
computer.
The mean square residual Lorentz forces of the reconstructed field
could be reduced to about a third of those of the discretized original
Low and Lou field, its divergence to 1/80 of
$\bra(\div{\vec{B}_{\mathrm{org}}})^2\ket$.

\subsection{Twisted loop}
\label{results:TwistedLoop}

As a second experiment we tried to model a twisted magnetic loop with the
$\alpha$ amplitude as a free parameter.
The boundary conditions for the GR iterations were chosen in the following
way:
\begin{gather}
\vec{n}\sdot\vec{B} = 0 \mathtext{and} \alpha = 0 
\mathtext{on} \partial V / \{\vec{x}\,|\,z=0\}\,,
\nonumber\\
\hat{\vec{z}}\sdot\vec{B} = \sum_{\sigma=\pm}
         \sigma\exp -\frac{(\vec{x}-\vec{x}_\sigma)^2}{r^2}
\mathtext{and} \alpha = \alpha_{\mathrm{max}} |\hat{\vec{z}}\sdot\vec{B}|
\mathtext{on} \partial V\!\cap\!\{\vec{x}\,|\,z=0\}\,,
\label{TwistedLoop}\end{gather}
where $\vec{x}_\sigma$ = $(0.5 - 0.2\sigma, 0, 0)^T$ and $r$ = 0.1.
Since $|\hat{\vec{z}}\sdot\vec{B}|$ has a maximum of 1, 
$\alpha_{\mathrm{max}}$ measures the maximum magnitude of $\alpha$ of
our boundary values.
We also rely on the symmetry of the boundary conditions and do not make the
distinction between $(\partial V)^{\pm}$ as is formally necessary
(see eq. \ref{BC}). Rather, we would like to see our weighting mechanism
(see section \ref{sec:GradRubin}) at work. All weights were set to
unity. The grid size was chosen as $n$ = 64.

\begin{figure}
  \hspace*{\fill}
\includegraphics[bb=80 180 530 600,clip,width=7.5cm,angle=-90]{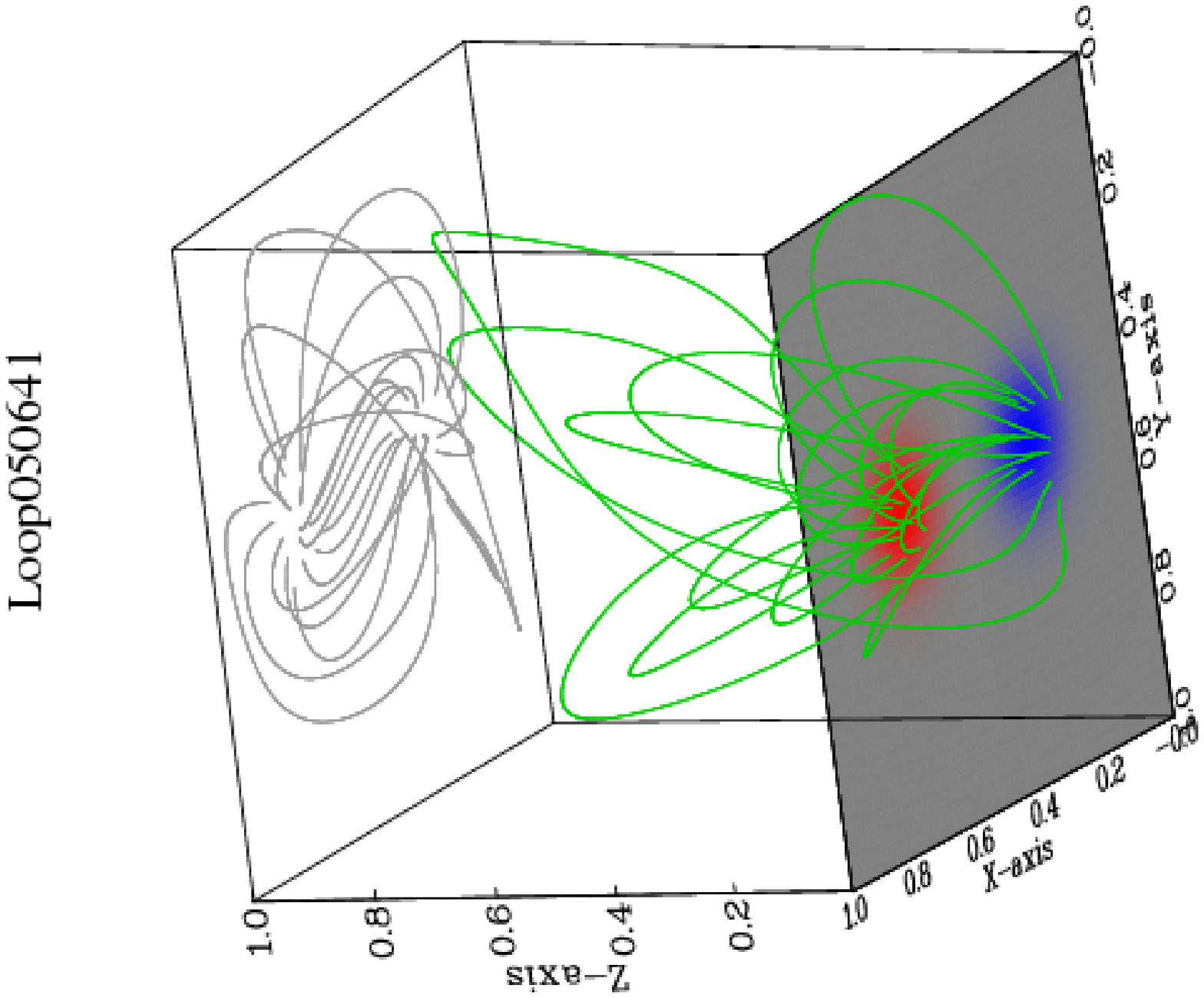}
\hspace{2mm}
\includegraphics[bb=80 180 530 600,clip,width=7.5cm,angle=-90]{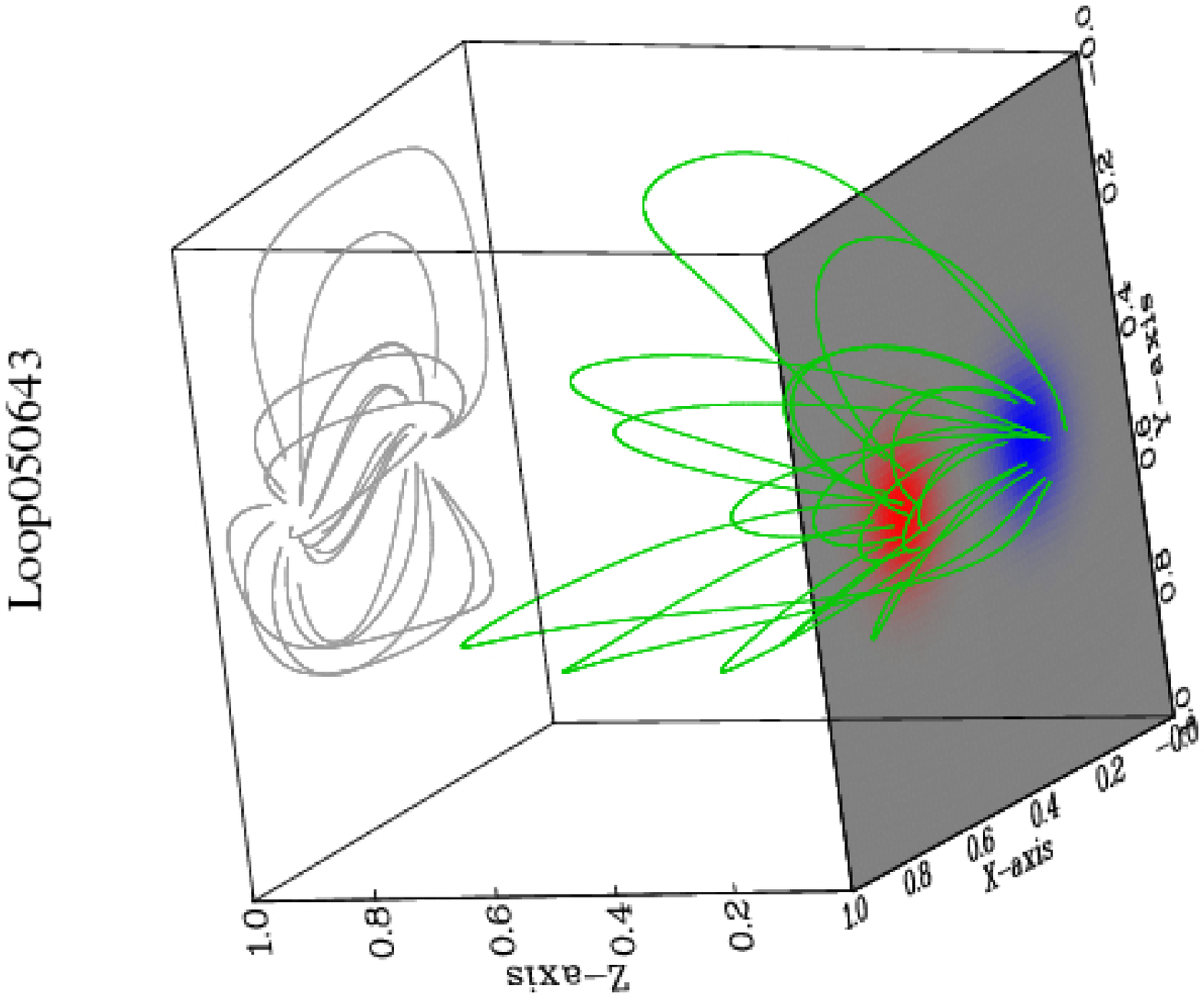}
  \hspace*{\fill}
  \caption{Representative field lines of the reconstructions of the
    twisted loop model by the GR ({\it left}) and WSR ({\it right})
    algorithms. In both cases $\alpha_{\mathrm{max}}$ = 5 was chosen.
    The {\it colour code} at the bottom represents $B_z$, the {\it top
    plane} shows the vertical projection of the field lines. The starting
    points of the field lines were chosen identically for both model results.}
\label{fig:PltFln2}\end{figure}

For the equivalent WSR calculations, we need to transform (\ref{TwistedLoop})
into the full field vector on the domain boundary. However only the
in-plane curl, $\vec{n}\crss\crl{\vec{B}}$ on $\partial V$ is determined
by the $\alpha$ boundary values.
We therefore use the boundary fields from the GR results as boundary
conditions for the WSR calculations. The local maximum discrepancy between
$\alpha$ as prescribed in (\ref{TwistedLoop}) and the GR result was less
than 10\%.

\begin{figure}
  \hspace*{\fill}
\includegraphics[bb=94 40 536 403,clip,width=8cm,angle=-90]{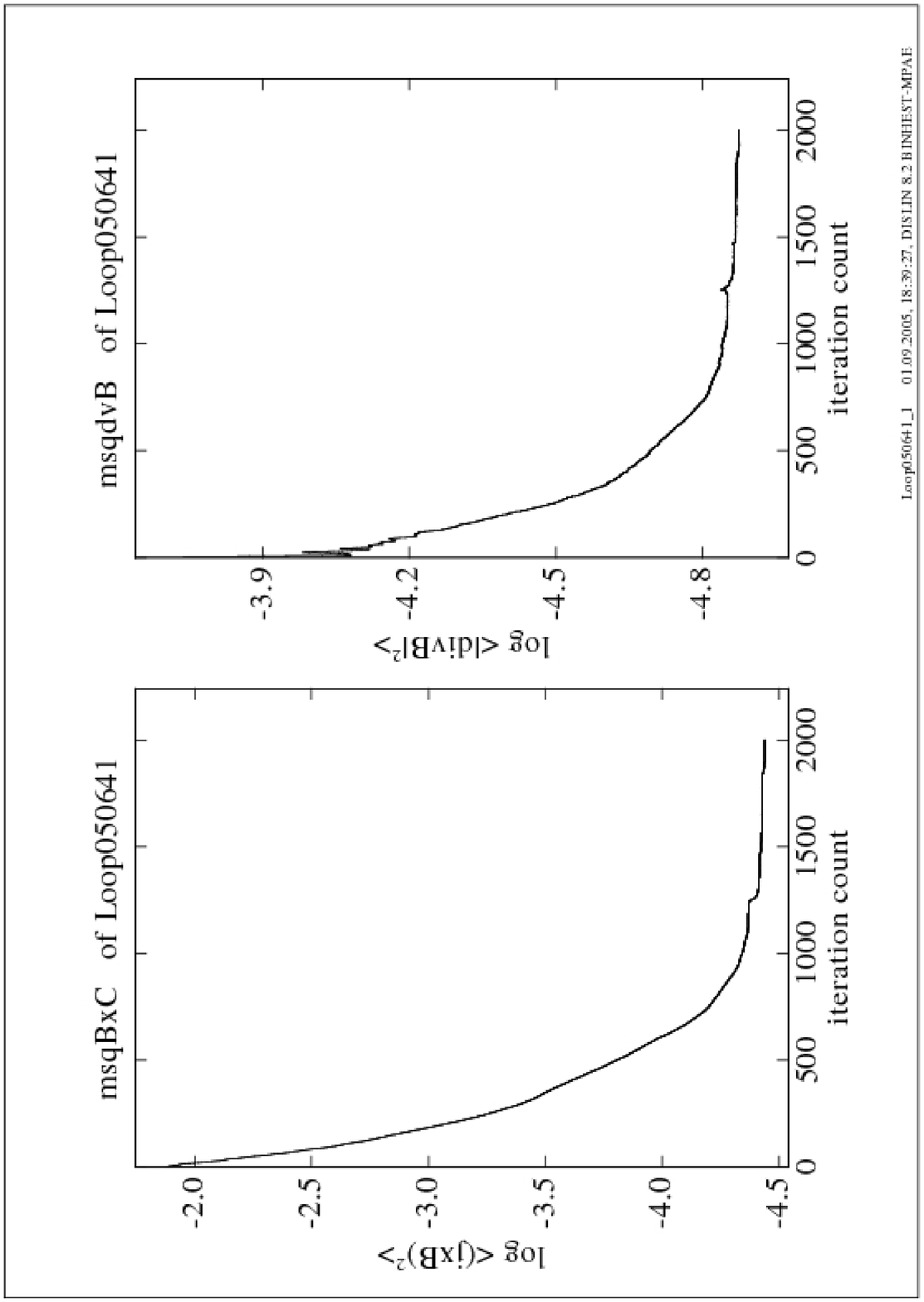}
\includegraphics[bb=94 40 536 403,clip,width=8cm,angle=-90]{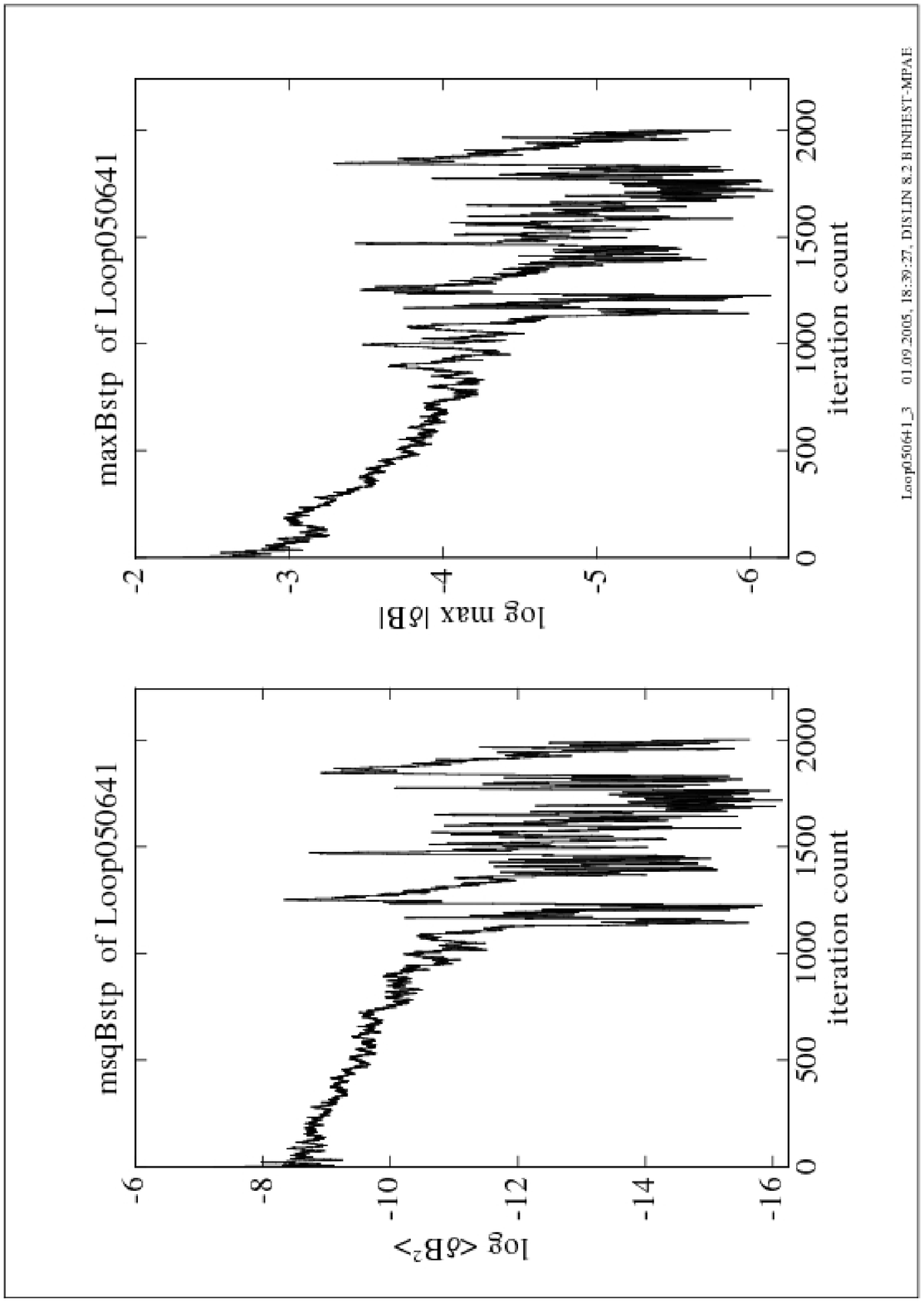}
  \hspace*{\fill}\\
  \caption{Mean square Lorentz force ({\it left}) and magnetic field
    correction step $\delta\vec{B}$ ({\it right}) for the twisted loop
    model reconstruction by the WSR code. The boundary conditions chosen
    correspond to $\alpha_{\mathrm{max}}$ = 5.}
\label{fig:WhLpHst}\end{figure}

In Figure~\ref{fig:PltFln2} we present the results of the different
algorithms for a value $\alpha_{\mathrm{max}}$ = 5. A visual
inspection reveals that the low lying loops correspond well in shape,
however the outer loops from the GR reconstruction show much more
twist than those produced with the WSR code.
The reasons for this difference may be twofold: The WSR result has the
larger residual Lorentz forces (see Figure~\ref{fig:WhLpHst}) and it
took almost 2000 iterations until a level was reached from which
$\bra(\vec{j}\crss\vec{B})^2\ket$ could not be lowered any further.
In the course of this WSR iteration we observed that the twist of the
outer field lines developed only very late. A continuation of the
iteration may therefore bring a slight improvement towards the
shape of the GR produced field lines.

The GR iterations, on the other hand, probably overtwist the outer loops.
The reason is the integration of $\alpha$ along field lines in (\ref{BgrdA}).
On the ground surface the vertical current density $\alpha B_z$ is confined
to only two small concentrated spots $\sim$ $\exp -2(\delta\vec{x}/r)^2$ of
less than 10 grid spacings in diameter. Even though we take great care
to follow as closely as possible the characteristics when we solve
(\ref{BgrdA}), a slight diffusion of $\alpha$ off the characteristics
can probably not be avoided. The diffused current density probably
causes a stronger twist for the outer field lines than is realistic.

The twisted loop model was designed to see how the codes behave when
$\alpha$ is enhanced. We pursued this question only for the GR code.
The amount of twist with which a field line can be charged, i.e., the
magnitude of $\alpha_{\mathrm{max}}$ is limited by the virial theorem
\citep{Molodensky:1969}.
The virial theorem states that for a sphere of radius $R$
\begin{equation}
   \int\limits_V |\vec{B}|^2 =
   R \int\limits_{\partial V} \big( |\vec{n}\sdot\vec{B}|^2
                                  - |\vec{n}\crss\vec{B}|^2 \big)\,.
\end{equation}
For a half space (i.e., $R$ $\rightarrow$ $\infty$) this reduces to
\begin{equation}
   \int\limits_{z=0} \big( |\hat{\vec{z}}\sdot\vec{B}|^2
                         - |\hat{\vec{z}}\crss\vec{B}|^2 \big)
                                \;\downarrow\; 0\,.
\label{AlyCrit}\end{equation}
This is in fact one of the relations enforced by the preprocessing scheme
for the boundary data by Wiegelmann et al. (\citeyear{Wiegelmann:etal:2005b}).
We can express the surface field in terms of a potential and a stream function
\[
\hat{\vec{z}}\crss\vec{B} = \hat{\vec{z}}\crss\grd{\phi} - \grd{\psi}\,,
\]
and it is clear that both $\phi$ and $\psi$ will depend on the shape
and amplitude of the $\alpha$ boundary condition at $z$ = 0. If the
boundary conditions are chosen as in (\ref{TwistedLoop}), then
\[
   \Delta\psi(\alpha) = \hat{\vec{z}}\sdot\vec{j}
                      = \alpha (\hat{\vec{z}}\sdot\vec{B})
                      = \alpha_{\mathrm{max}} |\hat{\vec{z}}\sdot\vec{B}|
                                              (\hat{\vec{z}}\sdot\vec{B})\,.
\]
The last step is due to the rigid connection between $\alpha$ and $B_z$ in
(\ref{TwistedLoop}).
Hence the stream function $\psi$ is directly proportional to
$\alpha_{\mathrm{max}}$ and invariable in its shape.
To keep the balance in (\ref{AlyCrit}), $|\grd{\phi}(\alpha_{\mathrm{max}})|$
has to decrease as $|\grd{\psi}(\alpha_{\mathrm{max}})|$ grows with
$\alpha_{\mathrm{max}}$.
Setting $\psi(\alpha_{\mathrm{max}})$ = $\alpha_{\mathrm{max}} \psi(1)$
and making use of the orthogonality (due to the symmetry in (\ref{TwistedLoop})
about $x$ = 0.5) of the two fields $\hat{\vec{z}}\crss\grd{\phi}$ and
$\grd{\psi}$ on the plane $z$ = 0 we obtain from (\ref{AlyCrit}):
\[
\alpha_{\mathrm{max}}^2 \le
       \frac{\int\limits_{z=0} \big(|B_z|^2
          - |\grd{\phi}(\alpha_{\mathrm{max}})|^2 \big)}
              {\int\limits_{z=0} |\grd{\psi}(1)|^2}
      \le
       \frac{\int\limits_{z=0} |B_z|^2}
              {\int\limits_{z=0} |\grd{\psi}(1)|^2}\,.
\]
A definite upper bound for $|\alpha_{\mathrm{max}}|$ is therefore
reached when $|\grd{\phi(\alpha_{\mathrm{max}})}|$ has declined to 0.
Numerically, we obtain for
(\ref{TwistedLoop}): $\int_{z=0} |B_z|^2$ = 0.031 and
$\int_{z=0} |\grd{\psi}(1)|^2$ = 0.00011 which gives
$|\alpha_{\mathrm{max}}|$ $\lesssim$ 16 as upper bound. Since
$|\grd{\phi(\alpha_{\mathrm{max}})}|^2$ probably never becomes zero, a
realistic upper bound is much less than this crude estimate.

\begin{figure}
  \hspace*{\fill}
\includegraphics[bb=97 40 531 403,clip,width=8cm,angle=-90]{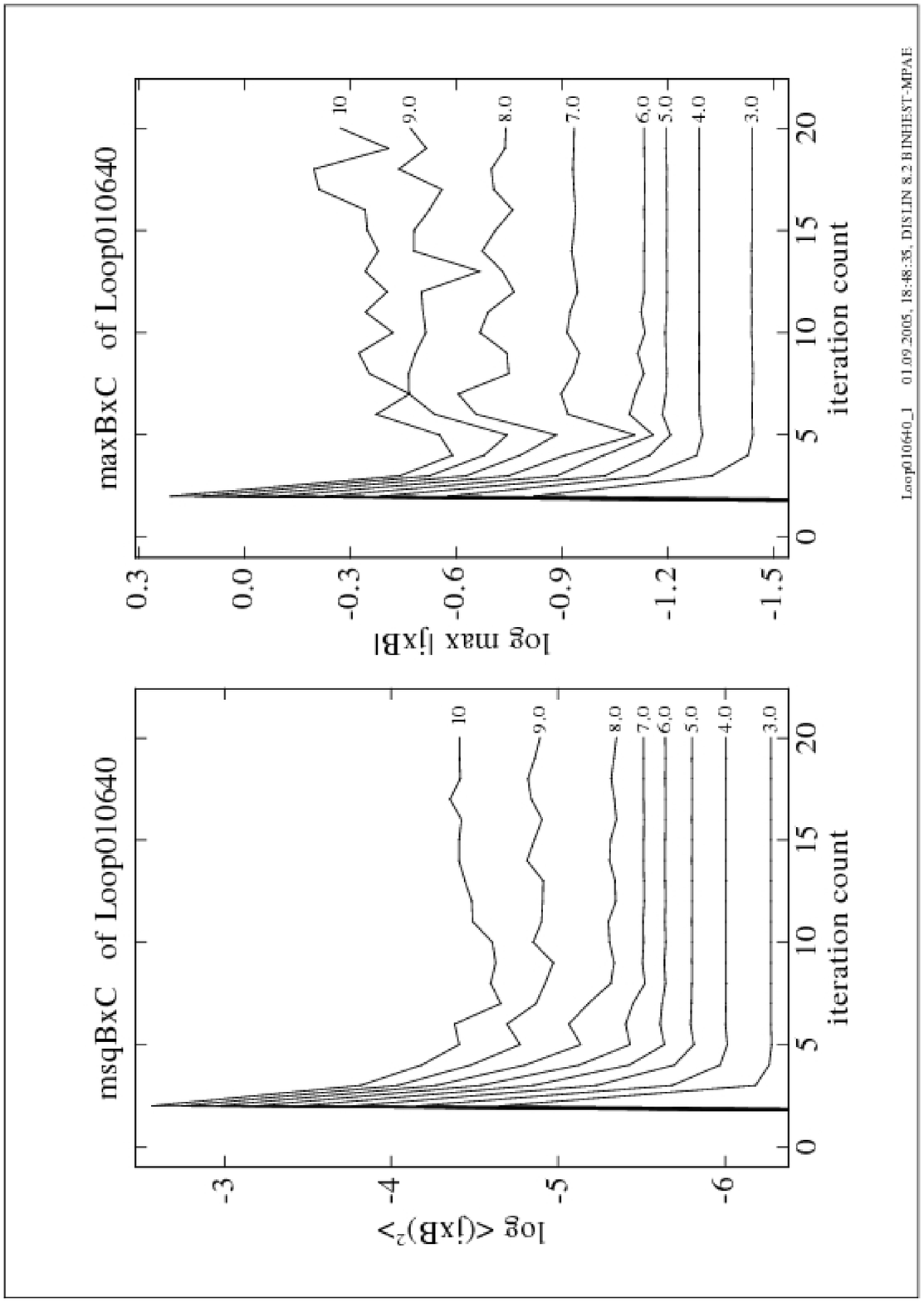}
\includegraphics[bb=97 40 531 403,clip,width=8cm,angle=-90]{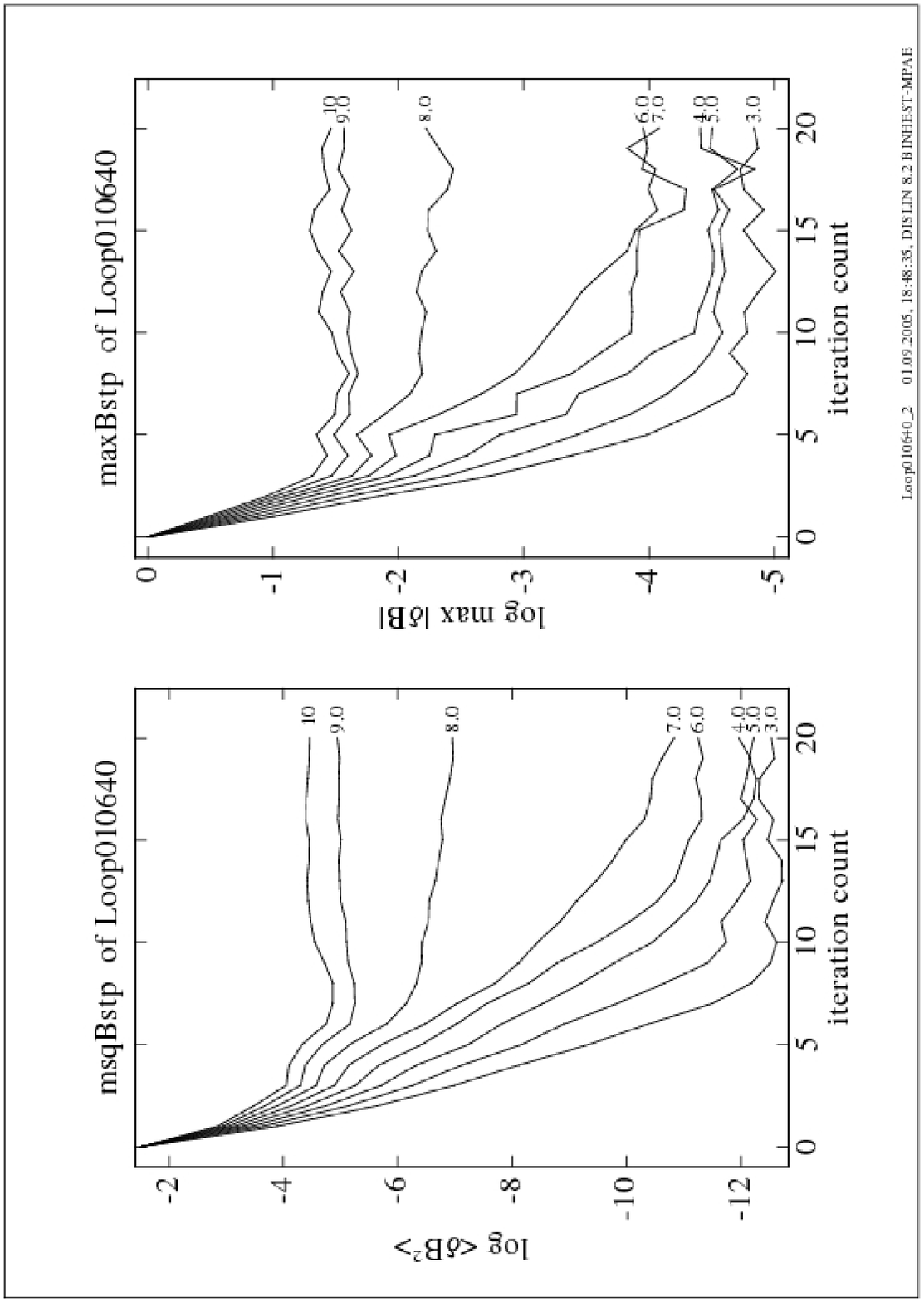}
  \hspace*{\fill}\\
  \caption{Mean square Lorentz force ({\it left}) and magnetic field
    correction step $\delta\vec{B}$ ({\it right}) for the twisted loop
    model with different amplitudes $\alpha_{\mathrm{max}}$ in the course
    of the GR iteration. The respective values of $\alpha_{\mathrm{max}}$
    are attached to the curve. Note the different ordinate scales compared
    to Figure~\ref{fig:WhLpHst}.}
\label{fig:GRLpHst}\end{figure}

In the GR iterations of the twisted loop model we varied
$|\alpha_{\mathrm{max}}|$ in the range [0,10]. We observe that the
convergence drastically decreases and comes to a halt between
$|\alpha_{\mathrm{max}}|$ = 7 and 8 (see Figure~\ref{fig:GRLpHst}). We
presume that this loss of convergence is not a failure of the code but
it rather indicates the upper limit of feasible values of
$|\alpha_{\mathrm{max}}|$. Beyond this limit, a stationary force-free
solution with a boundary as in (\ref{TwistedLoop}) probably is not possible.
Note that this transition is not so well visible in the residual
Lorentz forces which continually rise as $|\alpha_{\mathrm{max}}|$ and
hence the general level of the current density increases.

A point worthy of investigation is how the above estimate of the
maximum current density of a stationary force free flux tube complies
with the kink instability threshold for twisted flux tubes 
\citep[e.g.,][]{Mikic:etal:1990,VanHoven:etal:1995}. This instability
criterium predicts stationarity only if the number of turns a field makes
around the flux tube axis is less than about 2.4.

\section{Discussion}
\label{sec:Discussion}

Force-free extrapolation codes available nowadays are capable to
compute a field model in boxes as big as 256$\times$256$\times$256
\citep[e.g.,][]{Schrijver:etal:2006}. Compared to the resolution of modern
vector magnetograms with pixel arrays of 1000$\times$1000, the
calculated extrapolation models are most often quite limited in
their resolution.
Keeping in mind that they have to oversample the observation in order
to limit the discretization error, the observed data typically has to be
smoothed before it can serve as boundary condition for an extrapolation
\citep{Wiegelmann:etal:2005b}.
There is therefore a definite need for fast and effective force-free
reconstruction codes which can handle bigger models in order to make
proper use of the resolution which the observations provide.

In this paper we have implemented and tested two alternative schemes
for the extrapolation of a force-free field from boundary data.
The algorithms differ considerably and in order to compare these
different approaches unobscured by differences in the numerical
coding, we implemented them as far as possible in a similar way.
The discretization on a finite element grid by means of Whitney forms
results in codes with a very efficient performance. 
In addition, we have improved the WSR code considerably using an efficient
conjugate gradient iteration.

The two schemes differ not only in their approach towards a solution,
but they also differ in the boundary information which has to be supplied.
The WSR code requires boundary information which clearly overdetermines
the problem unless parts of the boundary fields are left to be varied.
But the full magnetic field vector on even part of the boundary is a very
strong constraint and there is a great danger to impose inconsistent
boundary values. This may be one reason why the WSR code converges
well for known solutions like the Low and Lou model \citep{Low:Lou:1990}
where precise and consistent boundary values are supplied, while a little
less reliable boundary values like those we retrieved from the result of
a GR extrapolation slow down the WSR convergence speed markedly. 

The WSR code has as a free parameter the weight between the
Lorentz force and the divergence term in (\ref{L_B}). In our
implementation it is hidden in the constant of the weight function $w$.
We found const $\sim$ 1 a good value for the problems which we have delt
with here. However, since the (\ref{FF}) is nonlinear, the performance
and also the optimal parameters may vary with the problem studied.
As an example, note the difference in the iteration history for
the Low and Lou model for const = 2 and 5 in Figure~\ref{fig:PltHstp}.
Here, the convergence changes drastically if the constant is modified
by only a small amount.

Typical computation times on a $n$ = 64 grid with our code are about
20 minutes on an ordinary home computer. At present, the codes still include
some checkout overhead, which could be dispensed with. Major improvements
can probably be obtained by spreading the calculation onto multiple grids.
We have made first tests with the WSR code for an adaptive enhancement of
the grid resolution. Here, the solution on the lower grid $n$ was used as
initial iterate on the next finer grid $2n$. We estimate that this way the
computation time can be reduced by about a factor 1/3.
Even more can probably be gained if the a true multigrid scheme is
used.

\section*{Acknowlegdements}
The work of Thomas Wiegelmann was supported by DLR grant 50 OC 0501.


\end{document}